\documentclass{mn2e}

\begin{document}

\title[Chandra Observations of NGC~4449]{Chandra Observation of NGC~4449. 
Analysis of the X-ray Emission from a Dwarf Starburst Galaxy.}

\author[L. K. Summers et al.]{Lesley K. Summers$^{1}$,
Ian R. Stevens$^{1}$, David K. Strickland$^{2}$\thanks{{\it Chandra}
Fellow} and \newauthor Timothy M. Heckman$^{2}$\\
  $^{1}$ School of Physics \& Astronomy, University of Birmingham,
Edgbaston, Birmingham, B15 2TT, UK\\
lks@star.sr.bham.ac.uk; irs@star.sr.bham.ac.uk\\
$^{2}$ Department of Physics \& Astronomy, The Johns Hopkins University,
3400 North Charles Street, Baltimore, MD 21218, USA\\
dks@pha.jhu.edu; heckman@pha.jhu.edu\\}

\date{Accepted .....................; Received .....................; 
in original form .......................}

\maketitle

\begin{abstract}
We present {\it CHANDRA} X-ray data of the nearby Magellanic Irregular
dwarf starburst galaxy NGC~4449. Contributions to the X-ray emission come
from discrete point sources and extended diffuse emission. The extended
emission has a complex morphology with an extent of $\sim 2.4\times
1.6$~kpc down to a flux density of $1.3 \times
10^{-13}$~erg~s$^{-1}$~cm$^{-2}$~arcmin$^{-2}$. The best spectral fit to
this emission is obtained with an absorbed, two temperature model giving
temperatures for the two gas components of $0.28 \pm 0.01$~keV and $0.86
\pm 0.04$~keV, a total mass content of $\sim 10^{7} M_{\odot}$ compared
with a galactic mass of several $10^{10} M_{\odot}$ and a total thermal
energy content of $\sim 2.5 \times 10^{55}$~erg, with an average energy
injection rate for the galaxy of a few $10^{41}$~erg~s$^{-1}$. Comparison
of the morphology of the diffuse X-ray emission with that of the observed
H$\alpha$ emission shows similarities in the two emissions. An expanding
super-bubble is suggested by the presence of diffuse X-ray emission within
what appears to be a cavity in the H$\alpha$ emission. The kinematics of
this bubble suggest an expansion velocity of $\sim 220$~km~s$^{-1}$ and a
mass injection rate of $\sim 0.14 M_{\odot}$~yr$^{-1}$, but the presence of
NGC~4449's huge HI halo ($r \sim 40$~kpc) may prevent the ejection, into the
inter-galactic medium (IGM), of the metal-enriched material and energy it
contains.

The arcsecond-resolution of {\it CHANDRA} has detected 24 X-ray point
sources down to a completeness level corresponding to a flux of $\sim 2
\times 10^{-14}$~erg~s$^{-1}$~cm$^{-2}$, within the optical extent of
NGC~4449 and analysis of their spectra has shown them to be from at least 3
different classes of object. As well as the known SNR in this galaxy, it
also harbours several X-ray binaries and super-soft sources. The point
source X-ray luminosity function, for the higher luminosity sources, has a
slope of $\sim -0.51$, comparable to those of other starburst galaxies.

\end{abstract}

\begin{keywords}
ISM: jets and outflows -- galaxies: individual: NGC~4449 -- galaxies:
starburst -- X-rays: galaxies.
\end{keywords}

\section{Introduction}
NGC~4449 is a nearby Magellanic Dwarf Irregular starburst galaxy which has
an inclination of 56.2$^{\circ}$ and a mass of $\sim 4 \times
10^{10}$~$M_{\odot}$ (Bajaja et al. 1994 - mass has been scaled for our
assumed distance which is discussed later). As such it allows another
opportunity for the study of the starburst phenomenon and its effect on
galaxy evolution in the local Universe.  Dwarf galaxies as the basic
building blocks in the hierarchical merging cosmology scenario are likely
to have harboured the earliest sites of star-formation in the Universe and
so their study in the local Universe can give insight into the evolution of
such objects at high redshift. Observations of local edge-on starburst
galaxies (Strickland et al. 2000; Weaver 2001) are presenting a picture of
kpc-scale, soft X-ray emitting, bi-polar outflows in the form of galactic
winds transporting mass, newly synthesized heavy elements and energy into
the IGM. These winds result from the pressure driven outflows along these
galaxy's minor axis produced from the efficient thermalization of the
mechanical energy from the supernovae (SN) explosions and stellar winds of
the massive stars in their OB associations and super star clusters (SSC).
Inclined galaxies such as NGC~4449 containing such phenomena present a less
clearly observable picture. However, observations of how absorption and
temperature of the diffuse X-ray emission varies within them and links
between H$\alpha$, HI and X-ray morphology suggest a similar scenario also
applies to these objects.

NGC~4449 has been observed across the electromagnetic spectrum and displays
some both interesting and unusual characteristics. Radio observations have
shown it to contain an extremely luminous SNR (Bignell \& Seaquist 1983),
have a very extended HI halo ($\sim 40$~kpc in radius) which appears to be
rotating in the opposite direction to the gas in the core of the galaxy
(Bajaja et al. 1994) and to have a large-scale ordered magnetic field
(Klein et al., 1996). The angular velocity of the rotation of the gas in
the centre of the galaxy is low ($\sim18$~km~s$^{-1}$ Hunter et al. 1998),
suggesting a low escape velocity but the huge HI halo may prevent escape of
the hot ejecta of the starburst region from the galaxy as a whole. The
galaxy has both a high star-formation rate ($\sim0.2$~M$_{\odot}$
~yr$^{-1}$, Thronson Jr. et al. 1987) and supernova rate, contains numerous
star clusters ($\sim60$, Gelatt et al. 2001) with the central one appearing
to be young ($\sim 6 - 10 $~Myr B\"oker et al. 2001) and a spherical
distribution of older stars with a mean age of $3-5$~Gyr (Bothun 1986). It
has also been shown to contain molecular cloud complexes from CO
observations (Hunter \& Thronson Jr. 1996) and has an infra-red luminosity
($10-150$~$\mu$m) of $3.7 \times 10^{42}$~erg~s$^{-1}$ (Thronson Jr. et
al. 1987, corrected for the distance assumed here - see below). The ionised
gas it contains shows a very disturbed morphology and includes what appear
to be many H$\alpha $ bubbles, shells and filaments (Hunter \& Gallagher
1990; 1997). The kinematics of the HII regions within the galaxy are very
chaotic and could be the aftermath of a collision or merger (Hartmann et
al. 1986). There is a gas-rich companion galaxy lying at a projected
distance of $\sim40$~kpc, coincident in projection with the outer edge of
the extended HI halo, which could have been involved. Previous X-ray
observations of this galaxy have detected the presence of discrete emission
from point sources and diffuse emission from hot gas (Della Ceca et
al. 1997; Vogler \& Pietsch 1997).

The distance to NGC~4449 is not well established and ranges from $2.93$~Mpc
(Karachentsev \& Drozdovsky 1998) to $5$~Mpc (Aaronson \& Mould 1983). We
have chosen to adopt the lower value throughout this analysis as optical
observations are showing objects within NGC~4449 to be resolved (Whiting
2002, private communication) suggesting a lower distance estimate to be
more appropriate.

In Section 2 we describe the {\it CHANDRA} observation. The X-ray emission
from the point sources is discussed in Section 3 and that from the diffuse
emission is considered in Section 4. Section 5 contains a more general
discussion of the relationship of the X-ray emission to that from other
wavebands, the affects of the out-flow of the X-ray emission on NGC~4449
and the morphology of the X-ray emission, whilst our main conclusions are
summarized in Section 6.

\begin{figure*}
\vspace{13cm} 
\includegraphics{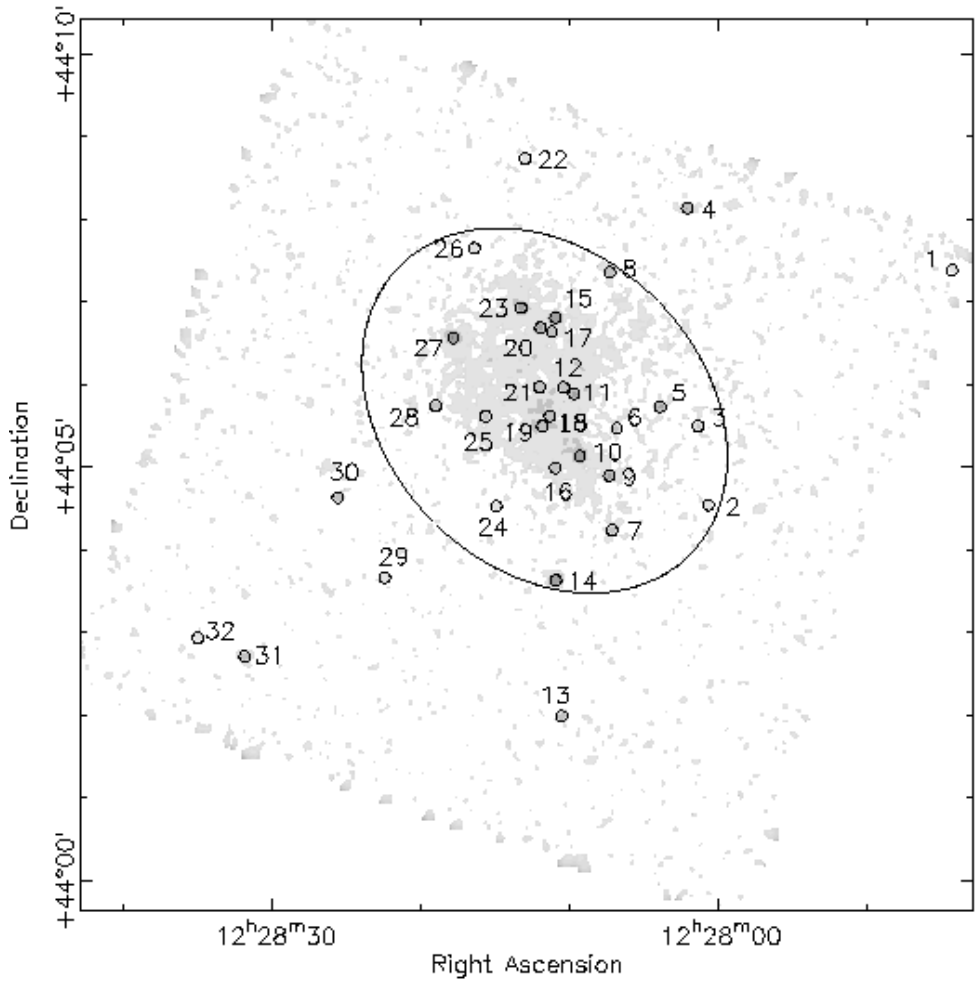}
\caption{Low resolution (smoothed using a Gaussian with FWHM of 4 pixels
$\sim2^{''}$), background subtracted image of the S3 chip field of view
marked with the 32 point sources listed in Table~1 and the $D_{25}$
ellipse, representing the optical extent of NGC~4449. The extended diffuse
emission associated with the galaxy is clearly visible.}
\label{sources}
\end{figure*}

\section{Observations and Analysis}

A 30~ks observation of NGC~4449 was obtained on Feb $4^{th} - 5^{th}$ 2001.
Analysis was carried out on the data contained within the S3 chip of the
ACIS$-$S instrument on board {\it CHANDRA} using {\it CIAO} (version
2.2.1), {\it HEASOFT} (version 5.1), {\it XSPEC} (version 11.1.0) and {\it
ASTERIX} (version 2.3-b1). The data were reprocessed (CALDB version 2.9)
and filtered to remove periods of flaring and lower than average count
rates using the $3\sigma$ clipping technique of the lc$\_$clean.sl script
(available from the {\it CHANDRA} website). After these processes were
completed, a total of 22.12~ks of useful data remained.  The data were then
further filtered to contain data in the energy band from $0.3 - 8.0$~keV
(amounting to a total of $\sim 22300$ counts on the S3 chip, of which $\sim
10700$ come from within the $D_{25}$ ellipse and of these $\sim 6100$ are
due to diffuse emission) and were also background subtracted using the
appropriately scaled CXC background event data set. (The one utilized for
this observation had a total exposure of 137391s.)  After production of an
exposure map, the point sources present in the data were detected using the
{\it wavdetect} tool, contained in the {\it CIAO} software, run with an
exposure map, default wavlet scales of 2.0 and 4.0 pixels and a source
significance threshold of $5 \times 10^{-7}$, a value of $\sim$~(number of
pixels)$^{-1}$ which should have limited the number of false detections to
$\sim 1$. 32 point sources were detected and these are shown in
Fig.~\ref{sources} overlaid on a smoothed image of the S3 chip (smoothed
with a Gaussian having a FWHM of 4 pixels $\sim2{''}$), while their
positions and background subtracted count rates are listed in Table~1. Of
these 32 sources, 8 do not lie within the optical extent of the galaxy, as
measured by the $D_{25}$ ellipse (de Vaucouleurs et al. 1991), also shown
in Fig.~\ref{sources}. The 24 sources within the optical extent of the
galaxy are the ones most likely to be associated with NGC~4449 and so are
the only ones considered further in this analysis. In comparison, when the
data were point source searched using {\it celldetect} with an exposure map
and default settings, only 18 sources were detected on the S3 chip. It was
obvious from visual inspection of the data that several significant sources
were missed by {\it celldetect}. In addition on comparison with expected
source counts from, `The {\it CHANDRA} Deep Field North Survey' (Brandt et
al. 2001), we would expect $\sim 16-20$ background sources on the S3-chip
at our level of completeness (see section 3.1 for further details). For
these reasons, we would expect more than the 18 sources detected by {\it
celldetect} and so the results from {\it wavdetect} have been adopted as
being more reliable.

\begin{table}
\begin{center}
\caption{Positions and count rates of the 32 sources detected in the
NGC~4449 {\it CHANDRA} S3 chip data. Column 1 is the source number ordered
in increasing R.A.. Columns 2 and 3 give the R.A. and Dec. of each source
and Column 4 lists their background subtracted count rates.}
\begin{tabular}{|c|c|c|c|} \hline
Source & RA (h m s) & Dec ($^{\circ} \; {'}\; {''}$) & Count Rate \\
 & & & ($\times 10^{-3}$ cts s$^{-1}$) \\ \hline
1 & 12 27 44.24 & 44 07 23.0 & $0.13\pm0.08$ \\
2 & 12 28 00.65 & 44 04 32.5 & $0.37\pm 0.14$ \\
3 & 12 28 01.36 & 44 05 29.9 & $0.63\pm 0.18$ \\
4 & 12 28 02.07 & 44 08 08.0 & $5.09\pm 0.48$\\
5 & 12 28 03.89 & 44 05 43.8 & $5.36\pm 0.50$\\
6 & 12 28 06.82 & 44 05 28.4 & $0.83\pm 0.21$\\
7 & 12 28 07.13 & 44 04 14.4 & $0.96\pm 0.22$\\
8 & 12 28 07.29 & 44 07 21.6 & $2.96\pm 0.37$\\
9 & 12 28 07.34 & 44 04 53.8 & $5.65\pm 0.52$\\
10 & 12 28 09.32 & 44 05 08.3 & $43.18\pm 1.42$\\
11 & 12 28 09.71 & 44 05 53.1 & $7.53\pm 0.59$\\
12 & 12 28 10.40 & 44 05 58.1 & $1.96\pm 0.32$\\
13 & 12 28 10.52 & 44 01 59.5 & $1.04\pm 0.24$\\
14 & 12 28 10.93 & 44 03 38.0 & $15.82\pm 0.87$\\
15 & 12 28 10.95 & 44 06 48.5 & $38.60\pm 1.33$\\
16 & 12 28 10.97 & 44 04 59.5 & $0.94\pm 0.23$\\
17 & 12 28 11.20 & 44 06 38.5 & $4.13\pm 0.44$\\
18 & 12 28 11.33 & 44 05 37.0 & $0.50\pm 0.19$\\
19 & 12 28 11.84 & 44 05 30.0 & $1.80\pm 0.32$\\
20 & 12 28 11.98 & 44 06 41.5 & $6.50\pm 0.56$\\
21 & 12 28 12.02 & 44 05 58.4 & $2.18\pm 0.33$\\
22 & 12 28 13.00 & 44 08 44.1 & $0.35\pm 0.13$\\
23 & 12 28 13.26 & 44 06 55.8 & $9.61\pm 0.68$\\
24 & 12 28 14.93 & 44 04 31.8 & $0.36\pm 0.14$\\
25 & 12 28 15.64 & 44 05 37.1 & $0.63\pm 0.19$\\
26 & 12 28 16.41 & 44 07 39.1 & $0.42\pm 0.14$\\
27 & 12 28 17.83 & 44 06 33.9 & $51.35\pm 1.55$\\
28 & 12 28 19.02 & 44 05 44.6 & $1.69\pm 0.28$\\
29 & 12 28 22.44 & 44 03 39.7 & $0.39\pm 0.14$\\
30 & 12 28 25.60 & 44 04 37.9 & $1.41\pm 0.26$\\
31 & 12 28 31.86 & 44 02 42.5 & $1.74\pm 0.33$\\
32 & 12 28 35.02 & 44 02 56.0 & $0.26\pm0.12$\\ \hline
\end{tabular}
\end{center}
\end{table}

\begin{figure*}
\vspace{11cm}  
\includegraphics{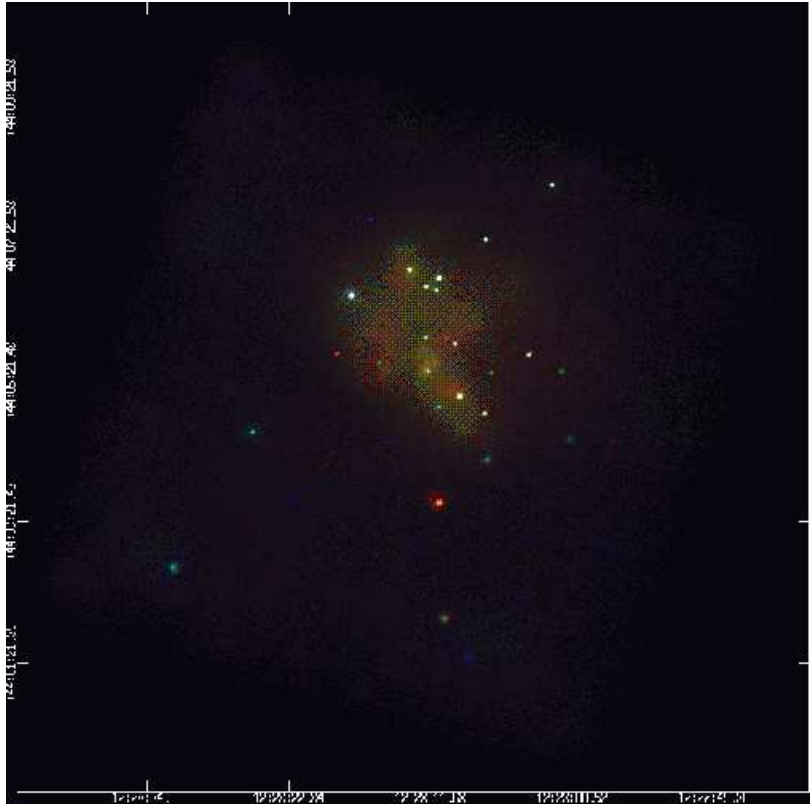}
\caption{Adaptively smoothed 3 colour image of NGC~4449 (red:
$0.3-0.8$~keV, green: $0.8-2.0$~keV and blue: $2.0-8.0$~keV). The extent of
the diffuse emission is $\sim 2.4$~kpc from NNE to SSW and $\sim 1.6$~kpc
from WNW to ESE. The variation of hardness in the spectra of the resolved
sources is evident from their varying colours and is indicative of the
presence of different types of sources, such as XRBs, SNRs and SSS.}
\label{3col}
\end{figure*}

After detection, the point sources were subtracted from the data so that
the diffuse X-ray emission contained within NGC~4449 could be
investigated. The sources were simply blanked, using areas equivalent to
the $3\sigma$ detection ellipses from {\it wavdetect} and allowance was
made for the lost flux from these areas, which amounted to only $0.5\%$ of
the total diffuse emission, in later analysis. Also, a 3 colour adaptively
smoothed image of the galaxy has been produced using {\it csmooth} in fft
(fast fourier transform) mode and this is shown in Fig.~\ref{3col}. The
lower and upper sigmas for deriving the smoothing kernel were set at 2 and
5 respectively. This image shows that some of the sources are very soft and
that the diffuse emission appears to consist of components of varying
temperatures. The red, green and blue images used correspond to energy
bands of $0.3-0.8$~keV, $0.8-2.0$~keV and $2.0-8.0$~keV respectively. These
bands were chosen so that the total number of counts in each band was
approximately equal thereby maximizing the S/N ratio for the 3 bands and
allowing point source searching of these individual bands to be performed
most effectively. These bands are also in line with those used in the
analysis of other starburst galaxies (e.g. NGC~253, Strickland et
al. 2002). The diffuse emission is seen to be extended and the size of the
region it occupies extends $2.4$~kpc from NNE to SSW and $1.6$~kpc from WNW
to ESE down to a flux density of $1.3 \times 10^{-13}
$~erg~s$^{-1}$~cm$^{-2}$~arcmin$^{-2}$.

\begin{figure}
\vspace{6.0cm}  
\includegraphics{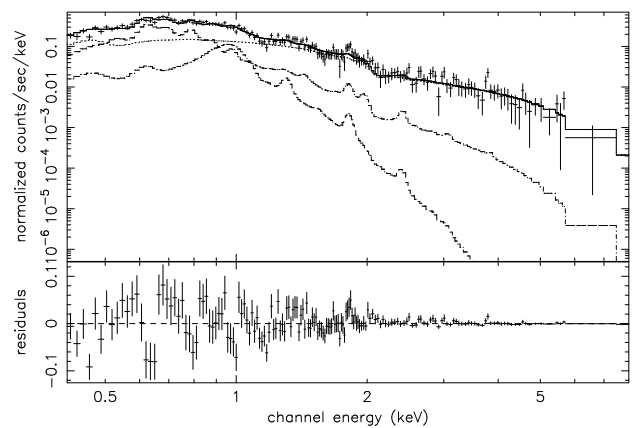}
\caption{Fitted spectrum of the total X-ray emission from NGC~4449. The fit shown
is an absorbed two temperature plus power law fit, with a column density of
$N_{H}=1.40 \times 10^{20}$~cm$^{-2}$ assumed for Galactic absorption and a
fitted column density of $N_{H}=(1.31 \pm 0.52) \times 10^{21}$~cm$^{-2}$
for the absorption within NGC~4449. The fitted temperatures of the two
thermal components are $0.27 \pm 0.01$ and $1.01 \pm 0.06$~keV with the
fitted abundance for both components being $(0.29 \pm^{0.54}_{0.08}
Z_{\odot}$ and the fitted photon index of the power law component having a
value of $\Gamma = 2.05 \pm 0.46$. The individual contributions of the
separate components of the fit are also included.}
\label{tot_spec}
\end{figure}

The background subtracted spectrum of the total X-ray emission from within
the $D_{25}$ ellipse of NGC~4449 was fitted using the modified
Levenberg-Marquardt method and standard $\chi^{2}$ statistic from {\it
XSPEC}, with an absorbed 2 thermal component plus power law fit. The
resulting spectral fit is shown in Fig. \ref{tot_spec}. The absorbing
column density was considered to be two components, one due to the
interstellar medium (ISM) within the Milky Way (Galactic column density,
$N_{H} = 1.4\times 10^{20}$~cm$^{-2}$) and the other due to the ISM of
NGC~4449. The 2 absorbed thermal components were modelled using the {\it
wabs} (photo-electric absorption using Wisconsin cross-sections) and {\it
mekal} thermal plasma codes within {\it XSPEC}, initially Galactic
absorption and an abundance of 0.276 Solar (the value obtained by Martin,
1997, for NGC~4449) were assumed. Two thermal plasmas were used to
represent the diffuse fraction of the total X-ray emission as images in the
soft ($0.3 - 0.8$~keV) and medium($0.8 - 2.0$~keV) energy bands, see
Fig. \ref{softmed}, show different spatial distributions of the hot gas and
also, multi-phase models are seen to be needed to best fit the emission
from other starburst galaxies (e.g. NGC~253, Strickland et al. 2002;
NGC~1569, Martin et al. 2002). A single temperature fit to NGC~4449 also
gave a less robust statistical fit and in particular when fitted to just
the diffuse emission does not have a wide enough energy distribution to fit
the spread seen in the data. The two-temperature model gives an absorption
corrected flux of $(2.39 \pm^{0.18}_{0.19}) \times 10^{-12}$~erg~s$^{-1}
$~cm$^{-2}$, which corresponds to a total X-ray luminosity, in the $0.3
-8.0$~keV band, of $(2.46 \pm^{0.19}_{0.20}) \times10^{39}$~erg~s$^{-1}
$. The fitted column density for NGC~4449 was $(1.31 \pm 0.52) \times
10^{21} $~cm$^{-2}$, with the two thermal components having temperatures of
$0.27 \pm 0.01$~keV and $1.01 \pm 0.06$~keV and a fitted abundance of $0.29
\pm^{0.54}_{0.08}$~Z$_{\odot}$. The power law component had a fitted photon
index of $\Gamma=2.05 \pm 0.46$. Of this total emission, $\sim 60\%$ is
from the resolved point sources, $\sim 30\%$ is from the cooler thermal
component and $\sim 10\%$ is from the hotter component.

\section{Point Sources}

\subsection{Source Spectra}

\begin{figure*}
\vspace{17cm}
 \includegraphics{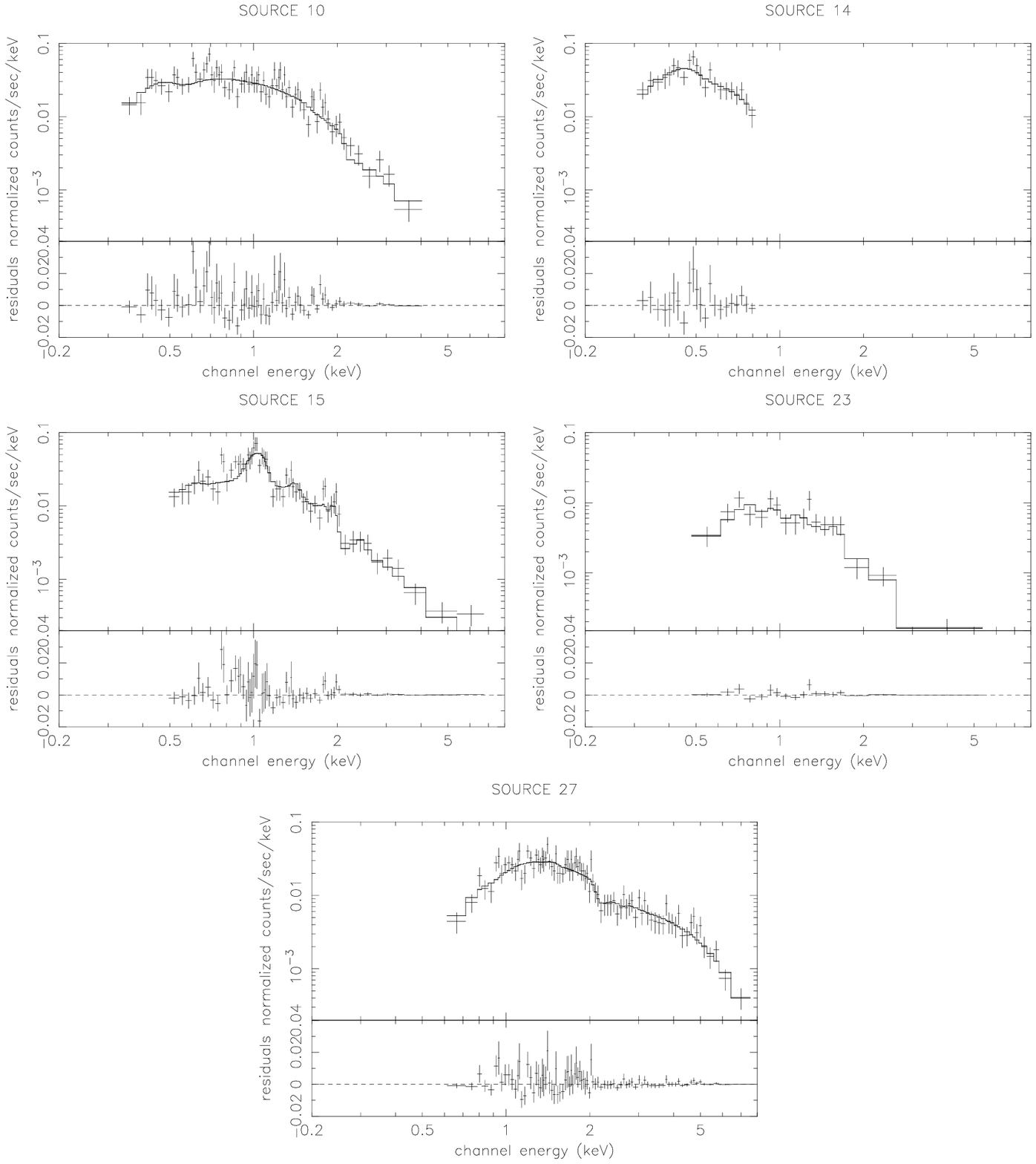}
\caption{Spectra and best fitting models for the brightest sources detected 
within the $D_{25}$ ellipse. Results are shown for source 10 (top left),
source 14 (top right), source 15 (middle left), source 23 (middle right)
and source 27 (bottom). All the fits are for the absorbed one component
models shown in Table~2. More detailed fits for the brighter sources are
discussed in Section 3.2 of the text.}
\label{spectra1}
\end{figure*}

A background subtracted spectrum was extracted for each of the 24 sources
within the $D_{25}$ ellipse. (The background spectra were taken from the
aforementioned background file and were appropriately scaled. The use of
the background files meant that the background spectrum was taken from the
identical position on the CCD and avoided the problem of contamination from
adjacent sources.) The spectra were grouped so that each had a minimum of 5
data bins and when possible a minimum of 10 counts per bin.  No attempt was
made initially to fit the data where a source had less than 50 counts after
background subtraction.  The fits obtained for the 5 sources with the
highest count rates are shown in Fig.\ref{spectra1}. Comparison of the low
count spectra to those fitted allowed a rough fit to be made to these
sources individually by using the models of sources with similar spectra
and hardness ratios (see below). The best fits obtained for the sources are
summarized in Table~2. The same start parameters were assumed for column
densities and abundances as were used for fitting the total emission.

\begin{figure}
\vspace{8.0cm}  
\includegraphics{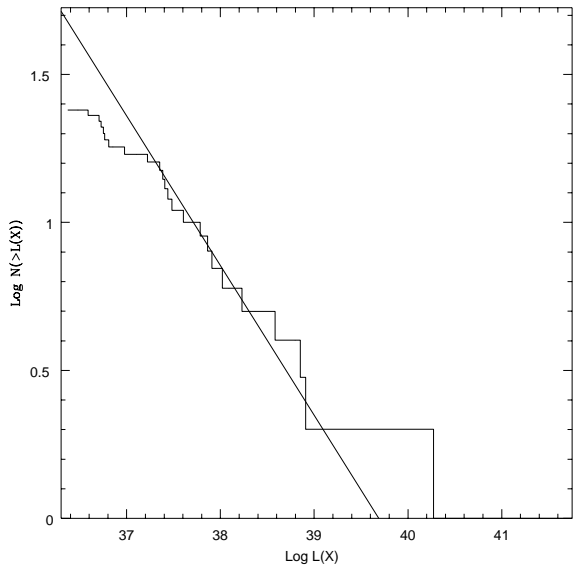} 
\caption{$\log(N)$ -- $\log(L_{X})$ plot for the 24 fitted resolved point sources
lying within the $D_{25}$ ellipse. The line shown is the power law fit to
the high luminosity end of the data. The slope of this fit is -0.51 and
incompleteness occurs at an absorption corrected luminosity of $2.18 \times
10^{37}$~erg~s$^{-1}$, which corresponds to a limiting absorption corrected
flux of $2.12 \times 10^{-14}$~ergs~$^{-1}$~cm$^{-2}$.}
\label{logn}
\end{figure}

For each of the individual source fits, the unabsorbed flux in the $0.3 -
8.0$~keV energy band and luminosity (assuming a distance of 2.93~Mpc for
NGC~4449) were calculated and these luminosities are also shown in Table~2
along with the errors determined from the $90\%$ confidence regions for
each model's normalization. In addition the data was point source searched
in the three different energy bands used to produce the three colour image,
(soft $0.3-0.8$~keV, medium $0.8-2.0$~keV and hard $2.0-8.0$~keV) so that
hardness ratios for the individual sources within the $D_{25}$ ellipse
could be calculated for the sources detected in more than one of the energy
bands. These ratios are shown in Table~3. The cumulative
Log(N)--Log(L$_{X}$) plot for the 24 fitted sources possibly associated
with NGC~4449 is shown in Fig.~\ref{logn}.  This has been fitted with a
single power law with a slope of $\sim-0.51$ for the higher luminosity
sources. The highest luminosity source is not included in the fit. To do so
results in an offset between the data and the fitted line but does not
alter the slope of the fit. The spectral fit to this source (source 28)
does suggest it is a very unusual source as it is very soft, lies behind a
high column density and has an extremely high luminosity
($>10^{41}$~erg~s$^{-1}$). The fact that its spectrum contains $<40$ counts
in total brings into question the robustness of the fit and for these
reasons it has been excluded from the fit. Sources with luminosities below
an absorption corrected luminosity of $2.18 \times 10^{37}$~erg~s$^{-1}$,
equivalent to a flux of $2.12 \times 10^{-14}$~erg~s$^{-1}$~cm$^{-2}$, were
also not included in the fit and these figures represent the completeness
limit for the data.  The fitted value for the slope is comparable to the
same fits for M82 and the Antennae where the slope was determined to be
$\sim-0.45\pm 0.06$ (Zezas et al. 2001).  The absorption corrected
contribution to the X-ray luminosity of NGC~4449 from unresolved point
sources is estimated to be $\sim (5.62 \pm ^{6.08}_{4.30}) \times
10^{37}$~erg~s$^{-1}$, equivalent to $\sim 5\%$ of the diffuse emission
(see section 4).

\begin{table*}
\begin{center}
\caption{Best fitting single component models for the spectra of the 24
point sources.  Column 1 gives the source numbers as shown on
Fig.~\ref{sources}. Column 2 contains the best fit absorbed single
component models obtained. In each case, the absorbing column due to the
Milky Way, wabs$_{Gal}$ is assumed to be $N_{H} = 1.4 \times
10^{20}$~cm$^{-2}$ Column 3 is the column density obtained from fitting the
wabs (photo-electric absorption using Wisconsin cross-sections) component,
column 4 the fitted temperature for the mekal, brems (thermal
bremsstrahlung), bb (black body), and nei (non-equilibrium ionization)
models for the thermal components, column 5 is the photon index for the po
(power law) models, column 6 the statistic of each fit and column 7 the
absorption corrected luminosity for each source. The errors shown on the
luminosities are from the $90\%$ confidence regions $(1.64\sigma )$ for
each model's normalization. Where only one value is shown this is an upper
limit from the same $90\%$ confidence regions. Column 8 lists the type of
source each object might be from comparison with the spectra and hardness
ratios of the identified sources within the galaxy.}
\begin{tabular}{|c|c|c|c|c|c|c|c|} \hline
Source & Model & $N_{H}$ (cm$^{-2}$) & $kT$ (keV) & $\Gamma$ & $\chi 
^{2}$ /d.o.f & Luminosity (erg~s$^{-1}$) & Source Type \\ \hline
2 & wabs$_{Gal}$(wabs(po)) & $3.14 \times 10^{21}$ & & 0.61 & 1.78/5 & $(7.10
\pm^{7.31})\times 10^{36}$ & AGN \\
3 & wabs$_{Gal}$(wabs(mekal)) & $3.68 \times 10^{21}$ & 0.90 & & 0.49/2 & $(5.87
\pm^{2.86}_{2.87}) \times 10^{36}$ & SNR/XRB \\
5 & wabs$_{Gal}$(wabs(po)) & $6.58 \times 10^{20}$ & & 1.30 & 1.39/8 & $(5.15
\pm^{0.75}_{0.92}) \times 10^{37}$ & XRB \\
6 & wabs$_{Gal}$(wabs(mekal)) & $1.31 \times 10^{21}$ & 0.82 & & 2.01/2 & $(5.51
\pm^{3.52}) \times 10^{36}$ & SNR/XRB \\
7 & wabs$_{Gal}$(wabs(po)) & $1.39 \times 10^{21}$ & & 0.40 & 1.00/2 & $(2.18
\pm^{0.98}) \times 10^{37}$ & AGN \\
8 & wabs$_{Gal}$(wabs(po)) & $1.53 \times 10^{20}$ & & 1.33 & 0.38/3 & $(2.59
\pm^{2.14}_{0.69}) \times 10^{37}$ & XRB \\
9 & wabs$_{Gal}$(wabs(po)) & $3.93 \times 10^{20}$ & & 2.88 & 7.87/8 & $(2.94
\pm^{0.66}_{0.56}) \times 10^{37}$ & SNR/XRB \\
10 & wabs$_{Gal}$(wabs(brems)) & $9.29 \times 10^{20}$ & 1.42 & & 83.37/69 & $(2.24
\pm^{0.12}_{0.12}) \times 10^{38}$ & SNR/XRB \\
11 & wabs$_{Gal}$(wabs(po)) & $3.50 \times 10^{21}$ & & 1.64 & 9.07/13 & $(8.91
\pm^{1.23}_{1.24}) \times 10^{37}$ & XRB \\
12 & wabs$_{Gal}$(wabs(bb)) & $4.15 \times 10^{20}$ & 0.091 & & 4.46/4 & $(1.28
\pm^{2.11}) \times 10^{37}$ & SSS \\
14 & wabs$_{Gal}$(wabs(bb)) & $8.96 \times 10^{20}$ & 0.097 & & 12.44/20 & $(1.24
\pm^{0.12}_{0.11}) \times 10^{38}$ & SSS \\
15 & wabs$_{Gal}$(wabs(nei)) & $1.26 \times 10^{21}$ & 2.58 & & 65.50/60 & $(2.34
\pm^{0.14}_{0.13}) \times 10^{38}$ & SNR \\
16 & wabs$_{Gal}$(wabs(po)) & $1.40 \times 10^{22}$ & & 2.53 & 0.90/3 & $(3.14
\pm^{1.16}_{1.23}) \times 10^{37}$ & XRB \\
17 & wabs$_{Gal}$(wabs(nei)) & $7.45 \times 10^{20}$ & 19.21 & & 5.33/5 & $(2.51
\pm^{0.50}_{0.49}) \times 10^{37}$ & SNR/XRB \\
18 & wabs$_{Gal}$(wabs(nei)) & $3.82 \times 10^{20}$ & 3.05 & & 1.03/3 & $(3.02
\pm^{1.37}) \times 10^{36}$ & SNR/XRB \\
19 & wabs$_{Gal}$(wabs(po)) & $1.31 \times 10^{21}$ & & 1.35 & 0.47/3 & $(2.33
\pm^{1.55}_{0.71}) \times 10^{37}$ & XRB \\
20 & wabs$_{Gal}$(wabs(po)) & $3.45 \times 10^{21}$ & & 2.76 & 8.20/10 & $(7.43
\pm^{1.04}_{1.23}) \times 10^{37}$ & SNR \\
21 & wabs$_{Gal}$(wabs(bb)) & $2.45 \times 10^{21}$ & 0.080 & & 1.71/2 & $(7.19
\pm^{1.07}_{2.76}) \times 10^{37}$ & SNR/SSS \\
23 & wabs$_{Gal}$(wabs(nei))  & $9.00 \times 10^{21}$  & 1.30 &  & 12.07/16  &
$(8.06 \pm^{0.98}_{1.01}) \times 10^{38}$ & SNR/XRB \\
24 & wabs$_{Gal}$(wabs(po)) & $1.10 \times 10^{22}$ & & 0.97 & 2.25/6 & $(5.79
\pm^{5.22}) \times 10^{36}$ & AGN \\
25 & wabs$_{Gal}$(wabs(bb)) & $1.11 \times 10^{21}$ & 0.130 & & 0.62/2 & $(5.17
\pm^{107.71}) \times 10^{36}$ & SSS/XRB \\
26 & wabs$_{Gal}$(wabs(brems)) & $4.72 \times 10^{21}$ & 54.85 & & 0.74/2 & $(4.99
\pm^{3.57}) \times 10^{36}$ & AGN \\
27 & wabs$_{Gal}$(wabs(po)) & $6.49 \times 10^{21}$ & & 1.91 & 64.38/86 & $(8.19
\pm^{0.42}_{0.41}) \times 10^{38}$ & XRB \\
28 & wabs$_{Gal}$(wabs(bb)) & $5.01 \times 10^{21}$ & 0.050 & & 1.28/3 & $(4.27
\pm^{1.32}_{1.34}) \times 10^{41}$ & SSS \\ \hline
\end{tabular}
\end{center}
\end{table*}

Earlier {\it ROSAT} HRI observations of this galaxy (Vogler \& Pietsch
1997) detected only 7 point sources within the $D_{25}$ ellipse. Of these
7, at least 3 of them have been resolved into more than one source using
the superior arcsecond spatial resolution of {\it CHANDRA} and an
additional 13 sources have also been detected. The source identified as the
supernova remnant (SNR), (X4) in the {\it ROSAT} data corresponds to source
15 in our {\it CHANDRA} data when the source position is compared to the
accurate position for the SNR obtained by Bignell \& Seaquist (1983), from
radio measurements and there are also 2 additional sources resolved at the
position of X4 in the {\it ROSAT} data. This highlights the fact that
luminosity estimates based on pre-{\it CHANDRA} data are effected by source
confusion.  Also of note is the fact that source 27 in our data ( X7 in the
{\it ROSAT} data) is much more luminous than it was in 1994 and now has the
highest count-rate for a single source in the galaxy rather than being
fourth highest. See Table~4 for a comparison of the $0.1-2.4$~keV
absorption corrected luminosities of sources that were both detected by
{\it ROSAT} and not resolved into more than one source by {\it CHANDRA}.

\begin{figure*}
\vspace{7cm}  
\includegraphics{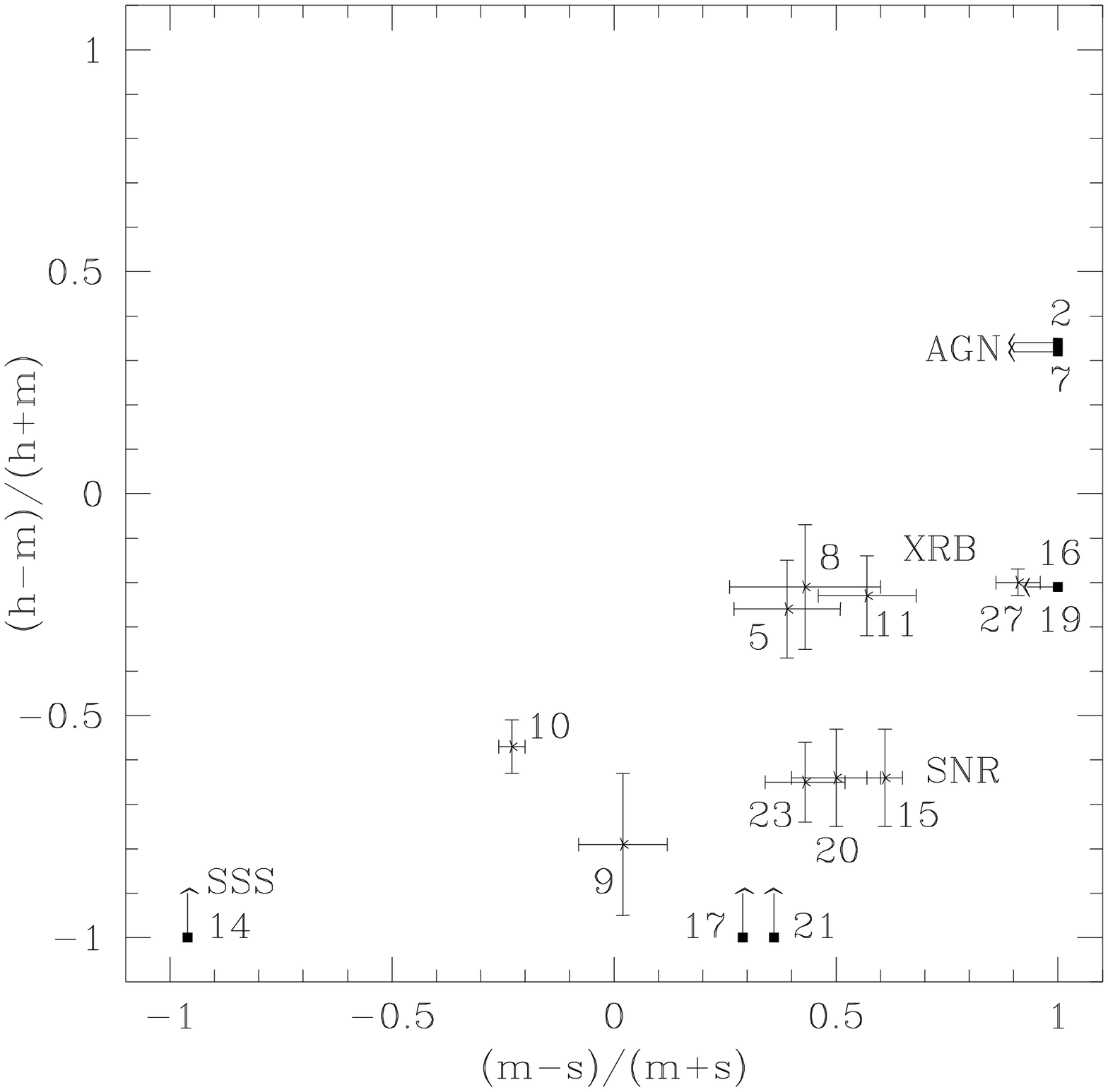} 
\includegraphics{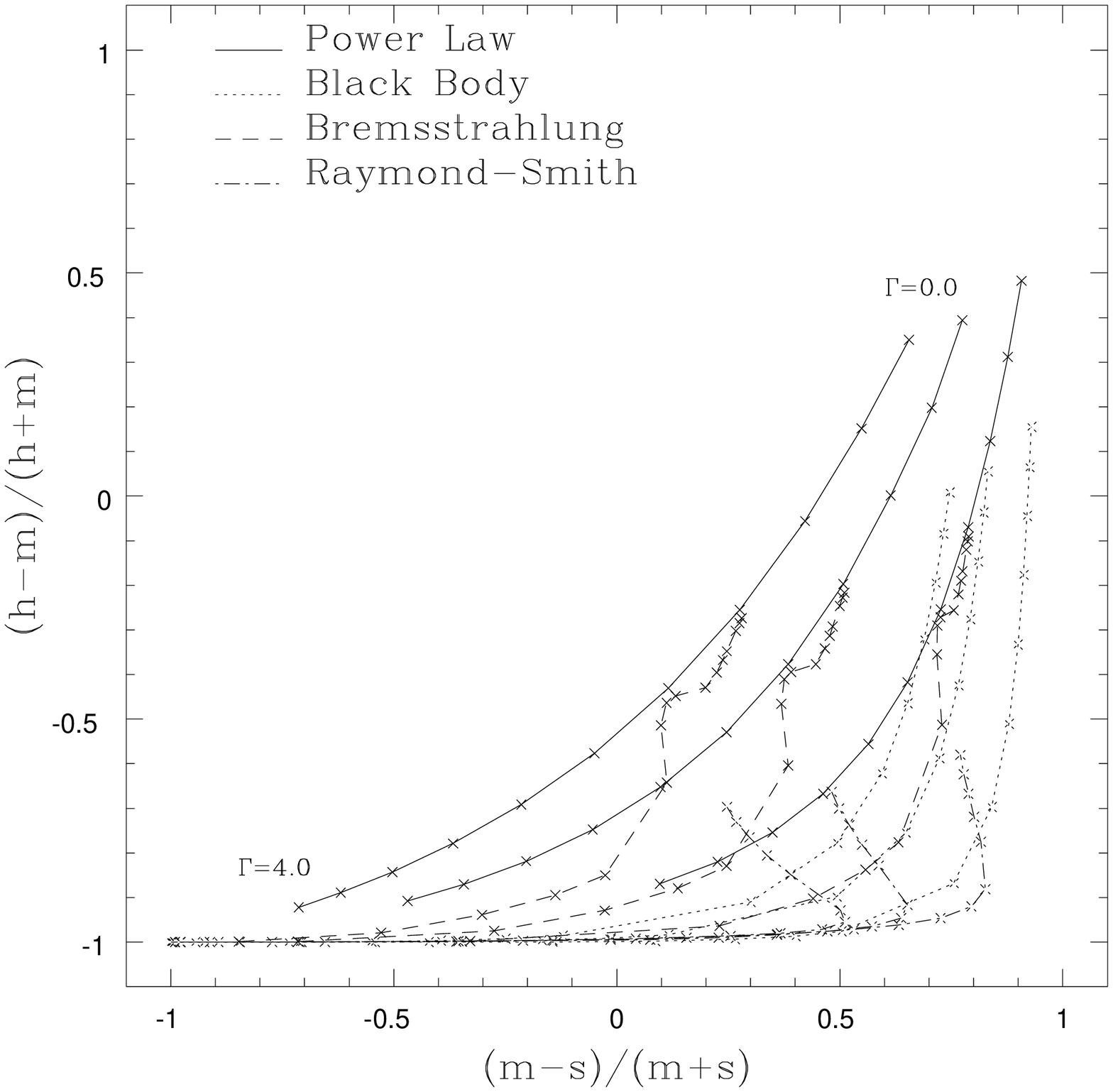} 
\caption{Left: Hard v. soft hardness ratios for the sources detected in the
separate energy bands. The hard hardness ratio is calculated from the
$2.0-8.0$~keV and $0.8-2.0$~keV energy bands and the soft hardness ratio is
calculated from the $0.8-2.0$~keV and $0.3-0.8$~keV energy bands in the
usual way as defined in the headings of columns 5 and 6 of Table~3. An
indication of the likely origin of the sources depending on the location in
this plot is also shown by the labels placed next to known sources. Right:
Theoretical tracks showing the positions different sources would occupy on
the hardness ratio plot. The three lines of each type have column densities
of $1.4\times 10^{20}$~cm$^{-2}$ (Galactic $N_{H}$), $1.0\times
10^{21}$~cm$^{-2}$ (fitted column density for the total emission from
NGC~4449) and $3.0\times 10^{21}$~cm$^{-2}$ (average fitted column density
for the fitted sources on the hardness ratio plot) respectively, moving
from left to right across the plot. For the three thermal models, the
temperatures increase from bottom left to top right. Black Body, $kT$
ranges from $0.04 - 1.0$~keV; Bremsstrahlung, $kT$ ranges from $0.2 -
100$~keV; Raymond-Smith, $kT$ ranges from $0.1 - 2.0$~keV and the plots
shown are for Solar abundance. Decreasing the abundance moves the
Raymond-Smith tracks to the left.}
\label{hr}
\end{figure*}

In general, the sources seem to split into four classes which can be
represented by sources 14, 15, 27 and 2. Sources like source 14 are very
soft, luminous sources (super-soft sources, SSS) which show little emission
above 1~keV. Source 15 is the SNR and at least 5 other sources show similar
spectra and/or hardness ratios. Sources like source 2 have spectra that
contain hard emission up to $\sim 5$~keV and sources in this group seem to
be best fitted by power law models. They are generally fitted with a higher
column density and lower photon index than other sources and may be
background AGN.  Sources like source 27 also have harder components to
their spectra, showing emission above 5~keV and are fitted by power laws
with $1.3 \leq \Gamma \leq 3.0$, typical of X-ray binaries (XRB). See
Section 3.2 below for further discussion on the brightest
sources. Fig.~\ref{hr} shows a comparison of the hard and soft hardness
ratios for the sources detected in all 3 energy bands. The hard ratio is
$(h-m)/(h+m)$ while the soft ratio is $(m-s)/(m+s)$, where $s$ is the count
rate in the soft band ($0.3 - 0.8$~keV), $m$ is the count rate in the
medium band ($0.8 - 2.0$~keV) and $h$ is the count rate in the hard band
($2.0 - 8.0$~keV). The sources represented by squares were only detected in
2 of the 3 energy bands and so their positions represent the most extreme
positions possible for them on this plot. Those at the bottom need to move
vertically upwards while those to the right will need to move horizontally
to the left.  The sources appear to clump in groups on this plot with the
SNR and SSS sources being softer than those we have classified as XRB and
AGN. The right hand plot in this figure shows some theoretical tracks of
how a sources position would vary depending on which type of spectral model
is assumed for it, what value the absorbing column density that it lies
behind has and what temperature or photon index is assumed for the model,
for comparison with the plotted source positions. An attempt has been made
to classify the sources using the shape and range of spectra and their
positions on the hardness ratio plot. The suggested classifications are
shown in column 8 of Table~2. The $D_{25}$ ellipse covers $\sim 25\%$ of
the S3-chip and as such it would be expected from the background number
counts quoted in section 2 that $\sim 4-5$ background sources should be
found within this area. This figure agrees well with the number of sources
classified as potential background AGN. In contrast to this, the 8 point
sources detected outside the $D_{25}$ ellipse are only about a half to
two-thirds of what would be expected, suggesting that our completeness
level estimate may be at slightly too low a flux.

\begin{table*}
\begin{center}
\caption{Hardness ratios, where possible, for the 24 sources detected
within the $D_{25}$ ellipse. Soft band, $s$, $0.3-0.8$~keV, medium band,
$m$, $0.8-2.0$~keV and hard band, $h$, $2.0-8.0$~keV. Where no counts are
shown, {\it wavdetect} failed to detect the object in that energy
band. Column 1 gives the source numbers as shown on
Fig.~\ref{sources}. Columns 2 -- 4 are the counts in the 3 different energy
bands and columns 5 and 6 give the values of the hardness ratios calculated
as detailed in the column headings.}
\begin{tabular}{|c|c|c|c|c|c|c|c|} \hline
Source & Counts in & Counts in & Counts in & \underline{(m - s)} &
\underline{(h - m)} \\ 
 & Soft Band & Medium Band & Hard Band & (m + s) & (h + m) \\ \hline
2 & - & $2.0 \pm 1.4$ & $3.9 \pm 2.0$ & - & $0.32\pm0.43$ \\
3 & - & $8.9 \pm 8.0$ & - & - & - \\
5 & $25.7 \pm 5.1$ & $58.7 \pm 7.7$ & $34.7 \pm 5.9$ & $0.39\pm0.12$ &
$-0.26\pm0.11$ \\
6 & - & $8.9 \pm 8.0$ & - & - & - \\
7 & - & $3.9 \pm 2.0$ & $7.9 \pm 2.8$ & - & $0.34\pm0.31$ \\
8 & $12.8 \pm 3.6$ & $31.8 \pm 5.7$ & $20.8 \pm 4.6$ & $0.43\pm0.17$ &
$-0.21\pm0.14$ \\
9 & $55.2 \pm 7.5$  & $57.4 \pm 7.6$ & $ 6.9 \pm 2.6$ & $0.02\pm0.10$ &
$-0.79\pm0.16$ \\
10 & $533.7 \pm 23.2$ & $336.9 \pm 18.4$ & $93.6 \pm 10.0$ &
$-0.23\pm0.03$ & $-0.57\pm0.06$ \\ 
11 & $23.9 \pm 5.0$ & $88.0 \pm 9.4$ & $54.6 \pm 7.4$ & $0.57\pm0.11$ &
$-0.23\pm0.09$ \\ 
12 & $39.6 \pm 6.4$ & - & - & - & - \\
14 & $349.0 \pm 18.7$ & $7.9 \pm 2.8$ & - & $-0.96\pm0.07$ & - \\
15 & $141.7 \pm 12.0$ & $590.5 \pm 24.4$ & $126.5 \pm 11.3 $ &
$0.61\pm0.04$ & $-0.64\pm0.11$ \\ 
16 & - & $10.5 \pm 3.3$ & $6.9 \pm 2.6$ & - & $-0.21\pm0.17$ \\
17 & $30.1 \pm 5.6$ & $54.1 \pm 7.4$ & - & $0.29\pm0.49$ & - \\
18 & - & - & - & - & - \\
19 & - & $22.4 \pm 4.9$ & $14.7 \pm 3.9$ & - & $-0.21\pm0.17$ \\
20 & $30.1 \pm 5.6$ & $91.1 \pm 9.6$ & $19.7 \pm 4.5$ & $0.50\pm0.10$ &
$-0.64\pm0.11$ \\
21 & $14.8 \pm 4.0$ & $31.7 \pm 5.7$ & - & $0.36\pm0.16$ & - \\
23 & $51.8 \pm 7.3$ & $129.6 \pm 11.4$ & $27.6 \pm 5.3$ & $0.43\pm0.09$ &
$-0.65\pm0.09$ \\
24 & - & - & $4.9 \pm 2.2$ & - & - \\
25 & - & - & - & - & - \\
26 & - & - & $4.9 \pm 2.2$ & - & - \\
27 & $32.4 \pm 5.7$ & $669.5 \pm 25.9$ & $448.7 \pm 21.2$ &
$0.91\pm0.05$ & $-0.20\pm0.03$ \\ 
28 & $36.6 \pm 6.1$ & - & - & - & - \\ \hline
\end{tabular}
\end{center}
\end{table*}

\begin{table*}
\begin{center}
\caption{Comparison of the absorption corrected luminosities for sources
10, 14, 23 and 27 obtained from {\it ROSAT} PSPC observations (11/1991),
{\it ROSAT} HRI observations (12/1994) and {\it CHANDRA} ACIS$-$S
observations (02/2001). The {\it ROSAT} values are from Vogler \& Pietsch
(1997) and are for 5~keV thermal bremsstrahlung spectra, corrected for
Galactic absorption. The first column of {\it CHANDRA} values are for the
best fits discussed in the text, corrected for Galactic absorption. All
luminosities are for the $0.1-2.4$~keV energy band and the {\it CHANDRA}
values have been scaled to account for the differences in assumed distances
to NGC~4449. The errors shown on the {\it CHANDRA} data are from the $90\%$
confidence levels ($1.64 \sigma$) of the normalizations of the
fits. Columns 1 and 2 are the source numbers allocated in the respective
data sets and columns 3 -- 5 give the absorption corrected luminosities
fitted to the 3 data sets. Column 6 shows the absorption corrected
luminosities obtained for the {\it CHANDRA} data when they are fitted with
a 5~keV thermal bremsstrahlung model, corrected for Galactic absorption.}
\begin{tabular}{|c|c|c|c|c|c|} \hline
Chandra & ROSAT & $L_{X}$(PSPC) & $L_{X}$(HRI) &
$L_{X}$(ACIS$-$S) & $L_{X}$(ACIS$-$S 5~keV) \\ 
 & & ($\times 10^{38}$~erg~s$^{-1}$) & ($\times 10^{38}$~erg~s$^{-1}$) &
($\times 10^{38}$~erg~s$^{-1}$) & ($\times 10^{38}$~erg~s$^{-1}$)  \\ \hline
10 & X1 & $2.92 \pm 0.43$ & $3.10 \pm 0.37$ & $3.24 \pm^{0.18}_{0.18}$  &
$2.58 \pm^{0.15}_{0.15}$ \\
14 & X3 & $0.70 \pm 0.23$ & $1.67 \pm 0.27$ & $105.45 \pm^{9.53}_{10.18 }$  &
$1.82 \pm^{0.20}_{0.21}$ \\
23 & X6 & $0.84 \pm 0.28$ & $0.76 \pm 0.21$ & $14.60 \pm^{1.87}_{1.68}$  &
$1.01 \pm^{0.12}_{0.13}$ \\
27 & X7 & $1.88 \pm 0.37$ & $1.00 \pm 0.21$ & $7.33 \pm^{0.38}_{0.36}$  &
$6.18 \pm^{0.33}_{0.33}$ \\ \hline 
\end{tabular}
\end{center}
\end{table*}

\subsection{More Detail on the Brightest Sources}

\subsubsection{Source 10 -- SNR?}
This source lies in the soft region of the hardness plot of Fig.~\ref{hr}
to the left of the area occupied by the known SNR. It is best fitted by an
absorbed thermal Bremsstrahlung model. Its position would also suggest it
should have a low column density as it lies to the left of the theoretical
tracks. This does not appear to be the case and fitting with other models
gives an even higher column density for a power law model or unphysical
results. This source was also detected (source X7) in the earlier {\it
ROSAT} observations and shows no evidence for variability as seen in
Table~4. Hence, it seems more likely that it is a SNR than an XRB.

\subsubsection{Source 14 -- SSS}
This source is very soft and it can be fitted in a similar way to CAL~87
(Ebisawa et al. 2001) using an absorbed blackbody component and several
absorption edges, indicative of a white dwarf binary system. For such a
fit, three absorption edges are included at energies of $0.40$~keV,
$0.59$~keV and $0.85$~keV the value of $N_{H}$ increases to $3.24 \times
10^{21}$~cm$^{-2}$ and $kT$ decreases to $0.065$~keV, compared to the one
component fit shown in Table~2, while the absorption corrected luminosity
in the $0.3-8.0$~keV band increases to $\sim 4.05 \times
10^{39}$~erg~s$^{-1}$. The absorption edges are possibly connected with the
presence of the following elements: $0.40$~keV, NVI or CVI; $0.59$~keV,
OVII or NVII; $0.85$~keV, OVIII and/or OVII (the low source counts making a
definite identification difficult).  The blackbody temperature fitted here
lies in the range ($T\sim 60 - 80$~eV) where both OVII and OVIII edges
would be expected to be present with the latter being more abundant
(Ebisawa et al. 2001). This source has a greatly increased absorption
corrected luminosity compared with the earlier {\it ROSAT} observations
however the difference is in part due to the different spectral models used
to fit the spectrum. The improved spectral resolution of {\it CHANDRA}
compared with {\it ROSAT} clearly shows that a $5$~keV thermal
bremsstrahlung model is inappropriate for this source, as all its emission
lies below 1~keV. In Table~4, the absorption corrected luminosity is also
much higher than that shown in Table~2 and quoted above due to the value in
Table~4 including emission down to $0.1$~keV rather than $0.3$~keV. These
facts make it difficult to assess if there has been any intrinsic increase
in the sources luminosity over the last decade. When the {\it CHANDRA} data
are fitted with the $5$~keV thermal Bremsstrahlung model there is agreement
with the {\it ROSAT} HRI data, within errors.

\subsubsection{Source 15 -- SNR}
This source has been identified as a young and very luminous SNR embedded
in an HII region from observations in several different wavebands, radio
(Seaquist \& Bignell 1978), optical (Balick \& Heckman 1978, Blair et
al. 1983), UV (Blair et al. 1984) and X-ray (Blair et al. 1983; Vogler \&
Pietsch 1997). Fitting this source with an absorbed, non-equilibrium
ionization model (as used for modelling other SNR, e.g. Yokogawa et al.,
2002) gives an X-ray temperature, $T_{X} \sim 2.2 \times 10^{7}$~K, a
column density of $N_{H}=1.26 \times 10^{21}$~cm$^{-2}$ and an absorption
corrected luminosity of $L_{X} = (2.34\pm^{0.14}_{0.13}) \times
10^{38}$~erg~s$^{-1}$ in the $0.3 - 8.0$~keV energy band, which are
comparable to the values recently obtained for the same SNR by Patnaude \&
Fesen (2003). (After allowing for differences in assumed distance, they
quote a luminosity of $L_{X} = 1.4 \times 10^{38}$~erg~s$^{-1}$ in the $0.5
- 2.1$ keV energy band, a temperature of $T \sim 9.2 \times 10^{6}$~K and
an absorbing column density of $N_{H}= 1.7 \times 10^{21}$~cm$^{-2}$.)
Following the same analysis that Blair et al. (1983) performed on this SNR,
we find firstly that our fitted temperature, is higher than the $6 \times
10^{6}$~K assumed by Blair et al. (1983, which came from the average
temperature determined from {\it Einstein} SSS observations of young SNR).
This higher temperature leads to an age for the SNR of $\sim 270$~yr and a
density of $120 - 200$~cm$^{-3}$ for the medium into which the SN
exploded. These results are higher than the age of $\sim 120$~yr and
density of $\sim 25$~cm$^{-3}$ reported by Blair et al. (1983). Allowing
for the difference in assumed distances to NGC~4449 would increase our age
estimate to $\sim 380$~yr without effecting the density. Compared to the
non-equilibrium ionization fit for the ionization timescale, where $\tau =
nt= 4.96 \times 10^{11}$~cm$^{-3}$~s, with the $90\%$ confidence regions
for this parameter giving a range of $3.61 \times 10^{11} \leq \tau \leq
5.00 \times 10^{13}$~cm$^{-3}$~s, the product of our calculated values
gives $1.0 \times 10^{12} \leq \tau \leq 1.7 \times 10^{12}$~cm$^{-3}$~s,
which gives reasonable agreement for the two methods.

\subsubsection{Source 23 -- XRB?}
The position of this source on the hardness plot of Fig.~\ref{hr} places it
in the group of XRBs however, its spectrum is best fitted by an absorbed
non-equilibrium ionization model and this source shows no strong evidence
for variability over the past 10 years, factors which may be more
indicative of a SNR than XRB.

\subsubsection{Source 27 -- XRB}
This source is now the source with the highest individual count rate in
NGC~4449. Its luminosity has increased by nearly an order of magnitude
since the {\it ROSAT} HRI observations of this galaxy in 1994, reported by
Vogler \& Pietsch (1997, see Table~4 for the comparative figures). This is
confirmed by the results obtained by fitting it with a $5$~keV thermal
bremsstrahlung model, as used for the {\it ROSAT} data, where its
luminosity is seen to have increased by a factor of at least 3. Its high
luminosity and apparent long term variability would suggest that it is an
high-mass X-ray binary (HMXB). The light curve shows no sign of variability
within our 30~ks observation.

\begin{figure}
\vspace{6.0cm}
\includegraphics{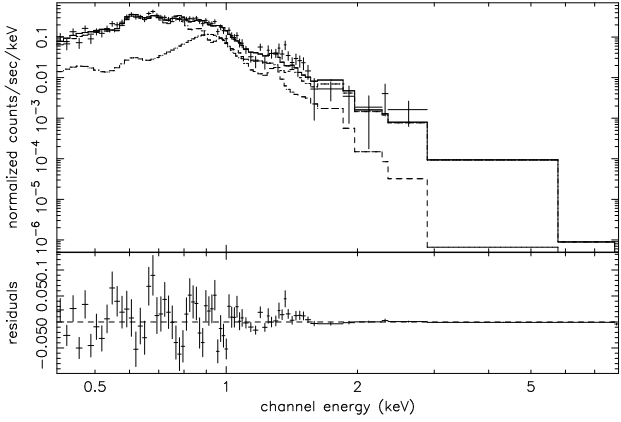}
\caption{Fitted spectrum of the NGC~4449 diffuse emission. The fit shown
is an absorbed two temperature fit, with a column density of $N_{H} = 1.4
\times 10^{20}$~cm$^{-2}$ assumed for Galactic absorption and fitted values
of $N_{H}=(1.29 \pm 0.38) \times 10^{21}$~cm$^{-2}$ for the absorption
within NGC~4449 and temperatures of $(0.28 \pm 0.01)$ and $(0.86 \pm
0.04)$~keV for the two thermal components. The fitted abundance for the two
thermal components was found to be $(0.32 \pm 0.08) Z_{\odot}$}
\label{diffspec}
\end{figure}

\begin{table}
\begin{center}
\caption{Parameters for the two gas components of the diffuse
emission. Assumptions: $V=6 \times 10^{65}$~cm$^{3}$ (assumes spherical
symmetry); $D=2.93$~Mpc; Filling factor, $f=1$; $n_{e} \sim(EI/Vf)^{1/2}$
where $EI$ is the emission integral (norm$\times 4\pi D^{2}$)/$10^{-14}$
and norm is the normalization obtained from the spectral fitting; $P\sim
2n_{e}kT$, $M\sim n_{e}m_{p}Vf$, $E_{Th} \sim 3n_{e}kTV$ and $t_{cool}
\sim (3kT)/(\Lambda n_{e})$ where $\Lambda = L_{X}/EI$.}
\begin{tabular}{|c|c|c|} \hline
Emission & Soft & Medium  \\ \hline
$kT$~(keV) & $0.28 \pm 0.01$ & $0.86 \pm 0.04$ \\
$T$~(K) & $(3.24 \pm 0.12) \times 10^{6}$ & $(9.97 \pm 0.46) \times 10^{6}$ \\
$L_{X}$~(erg~s$^{-1}$) & $(6.83 \pm^{0.35}_{0.36}) \times 10^{38}$ &
$(2.26 \pm^{0.27}_{0.26}) \times 10^{38}$ \\
$n_{e}$~(cm$^{-3}$) & $0.0122 \pm^{0.0003}_{0.0003}$ & $0.0059 \pm^{0.0004}_{0.0003}$ \\
$E_{Th}$~(erg) & $(9.66 \pm^{0.43}_{0.43}) \times 10^{54}$ &  $(1.49
\pm^{0.12}_{0.10}) \times 10^{55}$ \\
$M$~($M_{\odot}$) & $(6.02 \pm^{0.15}_{0.15}) \times 10^{6}$ &  $(3.01
\pm^{0.20}_{0.15}) \times 10^{6}$ \\
$P$~(dyn~cm$^{-2})$ & $(1.07 \pm^{0.05}_{0.05}) \times 10^{-11}$ &  $(1.65
\pm^{0.14}_{0.11}) \times 10^{-11}$ \\
$t_{cool}$~(yr) & $(4.66 \pm^{0.40}_{0.41}) \times 10^{8}$ & $(2.02
\pm^{0.39}_{0.36}) \times 10^{9}$ \\ \hline
\end{tabular}
\end{center}
\end{table}

\section{Diffuse Emission}

A point source and background subtracted spectrum of the diffuse emission
within the $D_{25}$ ellipse was extracted and this is shown in
Fig.~\ref{diffspec}. The fitted model shown is an absorbed 2 temperature
fit (wabs$_{Gal}$(wabs(mekal+mekal))). The component due to Galactic
absorption (wabs$_{Gal}$) has its absorption column density fixed at
$N_{H}= 1.4 \times 10^{20}$~cm$^{-2}$ and the fit gives $N_{H}= (1.29 \pm
0.38) \times 10^{21}$~cm$^{-2}$ for the absorbing column local to NGC~4449
and values of $kT$ of $(0.28 \pm 0.01)$~keV and $(0.86 \pm 0.04)$~keV for
the soft and medium thermal components respectively. The fitted abundance
for the two thermal components is $(0.32 \pm 0.08)Z_{\odot}$. The total
absorption corrected flux in the $0.3-8.0 $~keV energy band for the diffuse
emission is $(8.86 \pm^{0.61}_{0.60}) \times
10^{-13}$~erg~s$^{-1}$~cm$^{-2}$, which corresponds to an absorption
corrected luminosity of $(9.14 \pm^{0.63}_{0.62}) \times
10^{38}$~erg~s$^{-1}$.  Of this total emission, $75\%$ is in the soft
component and $25\%$ in the medium component and allowance has been made
for the flux lost during the point source subtraction.  The errors shown
here are based on the $90\%$ confidence levels $(1.64 \sigma )$ for the
normalization values obtained from the fit. Adding a power law component to
allow for unresolved point sources within the diffuse emission neither
improves the fit significantly nor alters the fitted parameters for the
thermal components. The photon index resulting from such a fit was $\Gamma
= 2.53$ and the fractional split of the flux between the 3 components was:
soft - $\sim 70\%$; medium - $\sim 25\%$; power law - $\sim 5\%$.

Other parameters of the two gas components have been calculated and these
are shown in Table~5. The assumptions of spherical symmetry and a filling
factor of 1 are both likely to be over estimates, particularly in the case
of the medium emission (see below), resulting in the figures being
under-estimates for $n_{e}$ and $P$ and over-estimates for $M$, $E_{Th}$
and $t_{cool}$ (see Strickland \& Stevens 2000 for a discussion of filling
factors and their likely values in galactic winds). These parameters will
be discussed further in section 5.2.
 
\begin{figure*}
\vspace{11.5cm}
\includegraphics{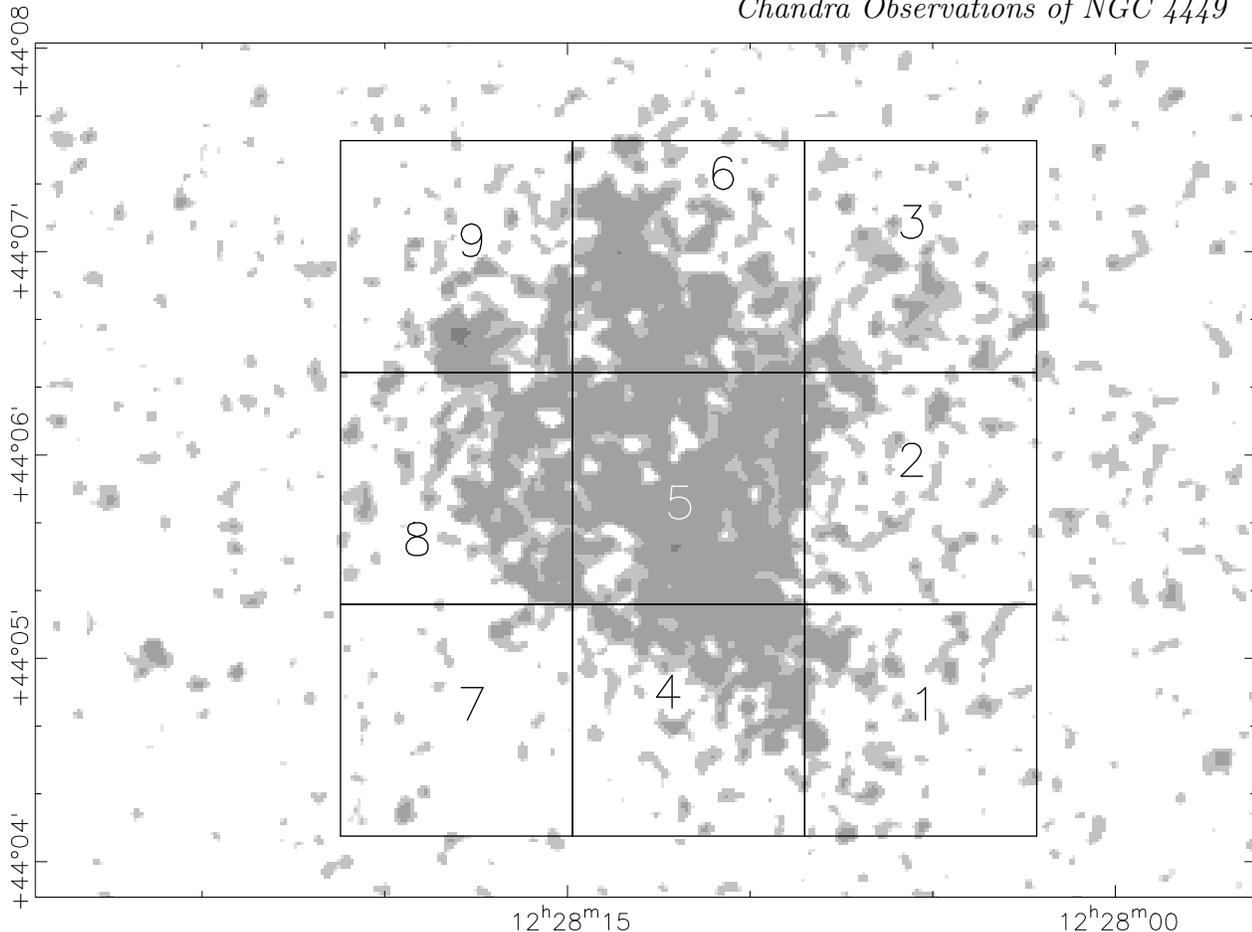} 
\caption{Low resolution (smoothed using a Gaussian with FWHM of 4 pixels
$\sim 2{''}$) image of the point source and background subtracted diffuse
X-ray emission overlaid with the 9 regions used to investigate the
variation of temperature and absorbing column within the diffuse
emission. The regions are identified by the overlaid numbers with region 1
in the SW corner and region 9 in the NE corner.}
\label{9}
\end{figure*}

In a further attempt to look at the abundances within the thermal
components and in particular the ratio of $\alpha$-elements to Fe the
spectrum was re-fitted using variable abundance mekal models for the two
thermal components. The abundances of the two thermal components were tied
and the individual abundances of Mg, Ne, Si and Ca relative to Solar values
were tied to that of O to form the group of $\alpha$-elements. The fitted
abundances for the $\alpha$-elements and Fe were $(0.27 \pm^{0.01}_{0.03})
Z_{\odot}$ and $(0.30 \pm^{0.02}_{0.02}) Z_{\odot}$ respectively, with a
$\chi^{2}$~/d.o.f of 201/117. The $\alpha/Fe$ ratio obtained from these
values is not significantly different from solar ($(0.91 \pm^{0.07}_{0.10})
Z_{\odot}$) in contrast to the values of Martin et al. (2002) for the dwarf
starburst NGC~1569, where values of $2.1 - 3.9 Z_{\odot}$ are quoted. The
difference in these values could reflect different contributions made to
the hot gas from SNe. Type II SNe are the sources of $\alpha$- elements
while Fe is produced from Type I SNe. The higher values for the $\alpha$/Fe
ratio for NGC~1569 would therefore suggest the presence of more Type II SNe
than in NGC~4449 and could suggest that NGC~1569 is experiencing shorter
more intense bursts of star-formation than NGC~4449 In the latter case, as
discussed by Della Ceca et al., 1997, the star-formation and energy
injection from star-forming regions may be more continuous as it is not
clear that the present star-formation rate is much larger than that in the
past on average and as such the longer time periods involved will reflect a
larger contribution from Type I SNe.

\begin{figure*}
\vspace{15.5cm}
\includegraphics{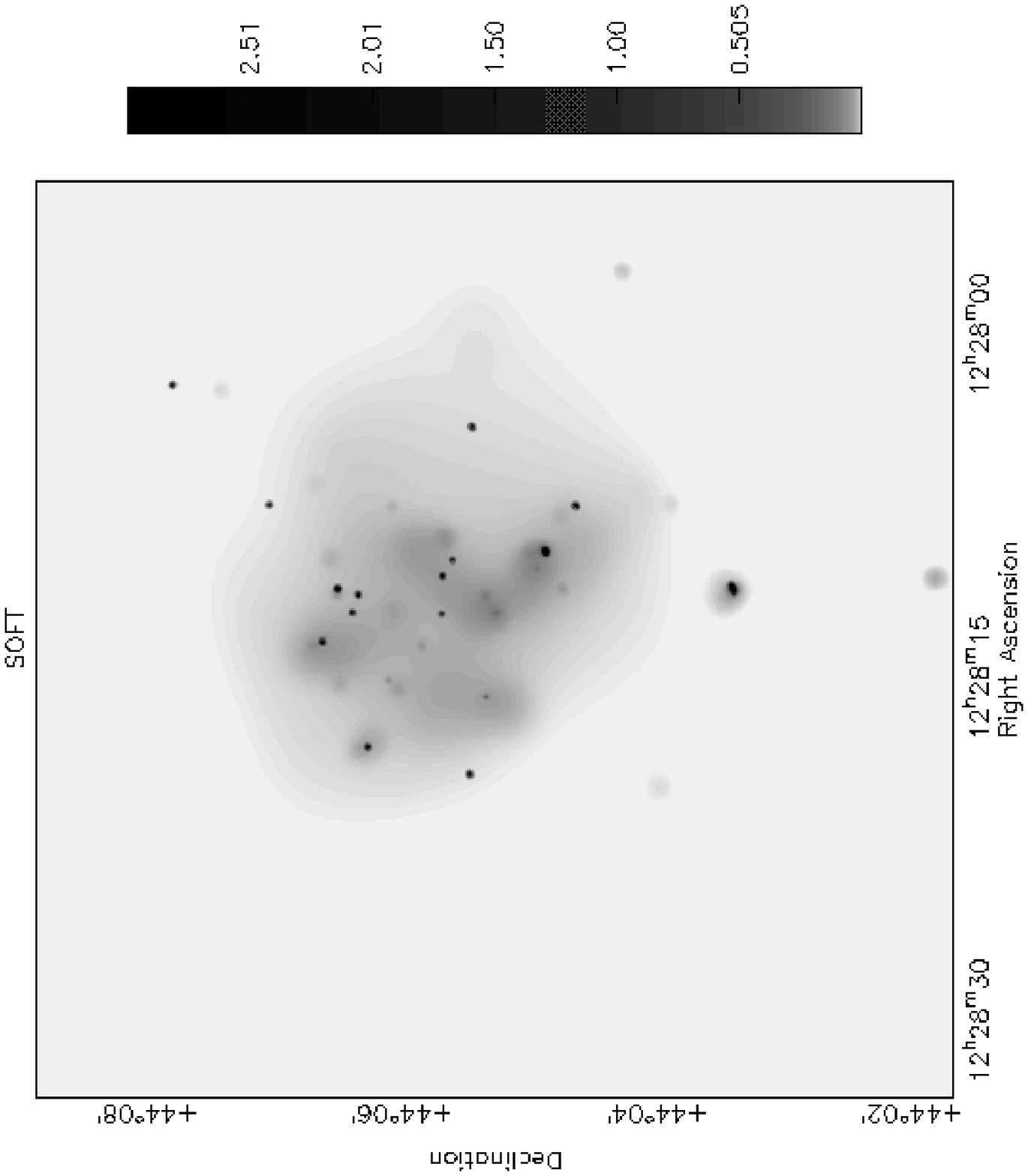}
\includegraphics{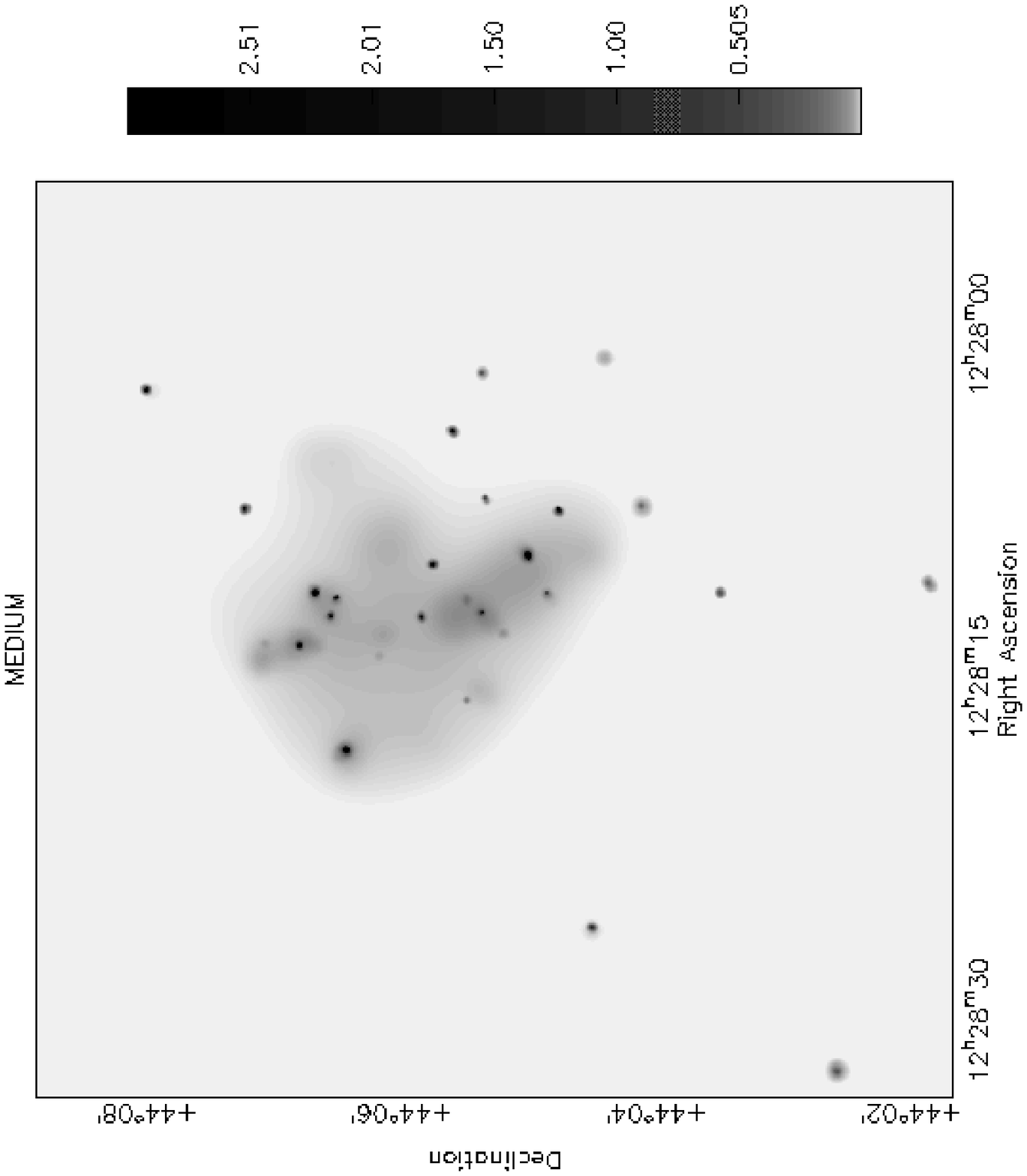}
\includegraphics{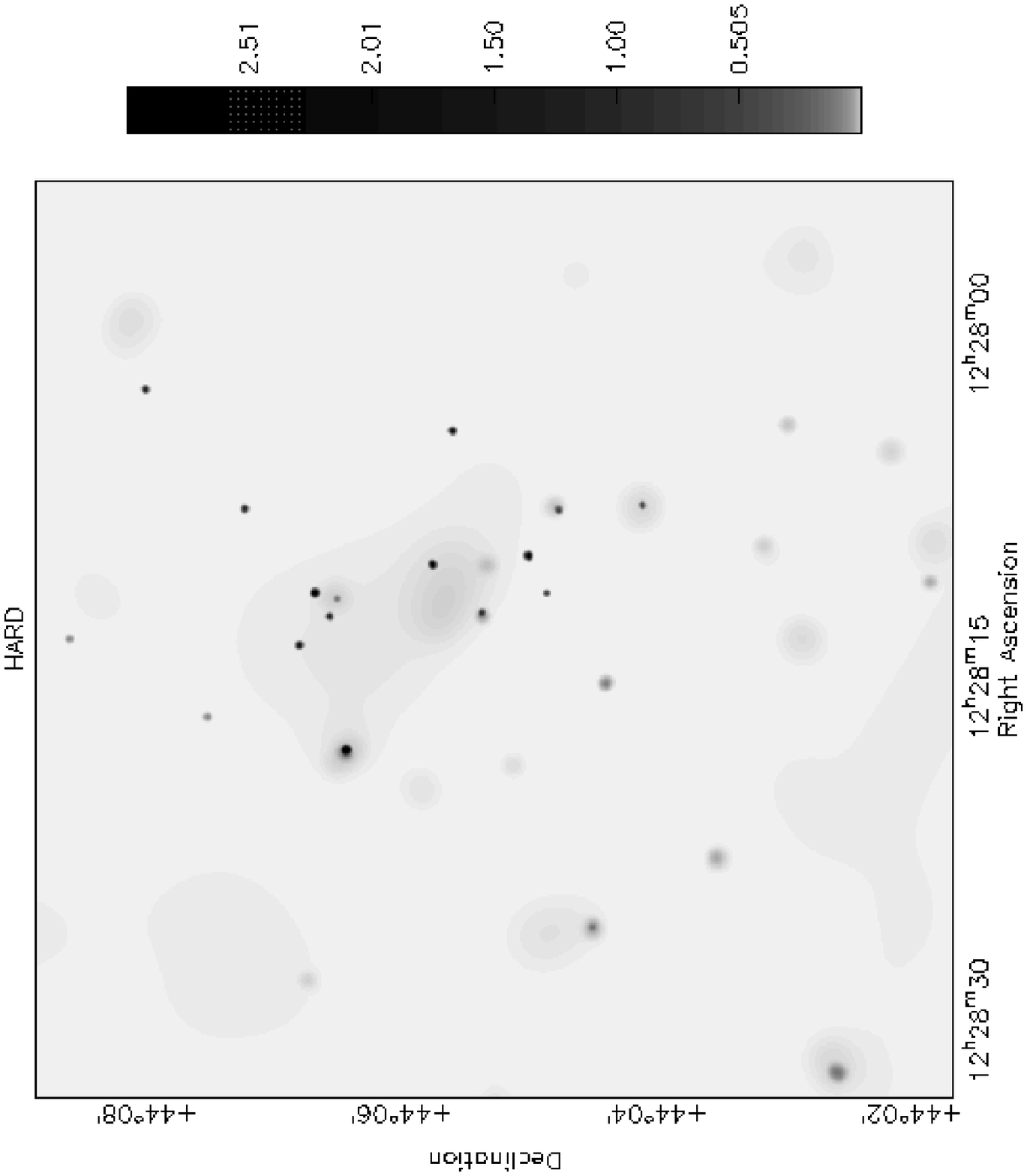}
\caption{The 3 panels show smoothed images of the soft
($0.3-0.8$~keV), medium ($0.8-2.0$~keV) and hard ($2.0-8.0$~keV) energy
bands, with the same grey-scale scaling, highlighting the differences in
spatial distribution between the bands for the diffuse emission and the
different spectral characteristics of the point sources.}
\label{softmed}
\end{figure*}

\begin{figure*}
\vspace{15.4cm}
\includegraphics{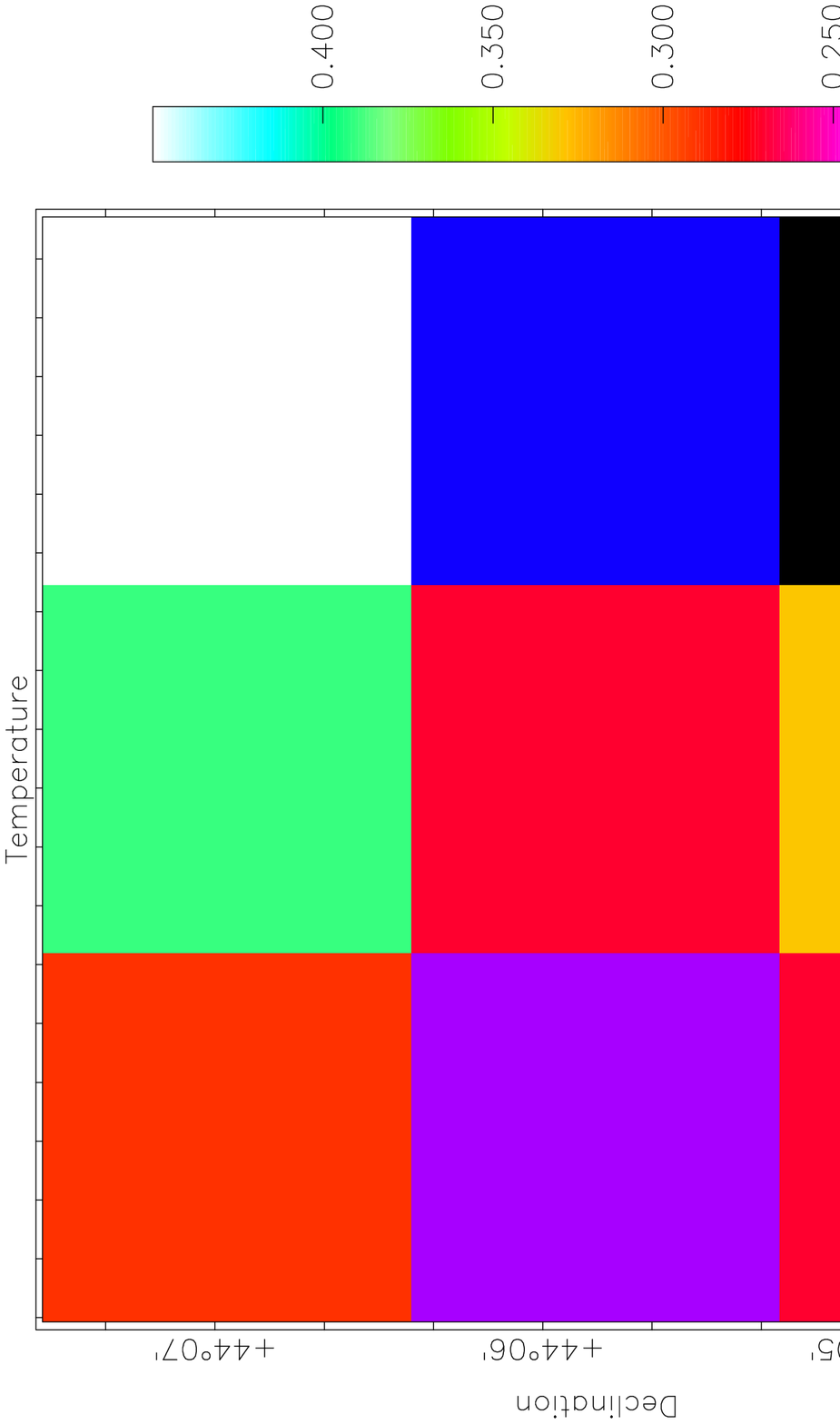}
\includegraphics{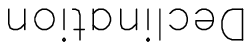}
\includegraphics{MC733_nhmap_fig10.ps}
\includegraphics{MC733_hrmap_fig10.ps}
\caption{Colour-coded maps of the parameters investigated for variation
within the diffuse emission. Top Left: The temperature. in keV, of the
softer thermal component within the 9 regions. Top Right: The temperature,
in keV, of the hotter thermal component, where two were fitted. For the
four regions with one temperature fits (regions 1, 2, 3 and 7), the soft
thermal component temperature is shown. Bottom Left: The logged value of
the fitted column densities (Log$(N_{H})$), in cm$^{-2}$, for the 9
regions. Bottom Right: The calculated $(m-s)/(m+s)$ hardness ratio for the
9 regions}
\label{maps}
\end{figure*}

In order to look for variations in temperature within the diffuse emission,
the region it occupies was divided into 9 regions and the spectra for each
region was extracted and fitted.  The regions are shown overlaid on the
smoothed (using a Gaussian with FWHM of 4 pixels $\sim 2{''}$), point
source and background subtracted image of NGC~4449 in Fig.~\ref{9} and the
resulting spectral fits are shown in Table~6. Where possible, a 2
temperature fit was used, but for the 4 regions with the lowest number of
counts (1, 2, 3 and 7) this was not possible. Either the fitted
temperatures of the 2 components were the same, within errors, or the
normalization for one of the components went to zero. For these 4 regions,
the results of single temperature fits are given.  In addition, for each
region, the $(m-s)/(m+s)$ hardness ratio has been calculated to further
highlight variations within the galaxy and these are shown in Table~7. To
make a comparison of the results contained in Tables 6 and 7 clearer,
colour-coded maps of the temperatures, column densities and hardness ratios
are shown in Fig.~\ref{maps}. The two upper panels compare the fitted
temperatures for the 9 regions. The one on the left shows the temperatures
of the softer thermal components, where two temperatures were fitted and
the right panel shows the temperatures of the harder components. Where only
one temperature was fitted, this is shown in both panels. The lower left
panel compares the column densities (Log$(N_{H})$) fitted to the 9 regions,
while the lower right one shows how the $(m-s)/(m+s)$ hardness ratio
varies.

\begin{table*}
\begin{center}
\caption{Single or two temperature fits for the 9 regions of the diffuse
emission (see text for further details). Columns 1 and 2 contain the region
numbers and counts for each region respectively. Column 3 lists the column
density for the absorbing gas within NGC~4449, column 4 the temperature(s)
and column 5 the statistic for the fit for each region. Column 6 has the
calculated absorption corrected luminosities for each of the regions, the
errors shown are from the $90\%$ confidence ranges $(1.64 \sigma )$of the
normalizations of the fits or are upper limits determined from these values
where only one value is given. In all cases, the Galactic absorption column
density was fixed at $N_{H} = 1.4 \times 10^{20}$~cm$^{-3}$.}
\begin{tabular}{|c|c|c|c|c|c|} \hline
Region & Counts & $N_{H}$~ (cm$^{-2}$) & $kT$~ (keV) & $\chi^{2}/$~d.o.f &
$L_{X}$~ (erg~s$^{-1}$)  \\ \hline 
1 & $197 \pm 21$ & $2.87 \times 10^{21}$ & 0.19 & 17.9/14 &
$(2.29 \pm^{0.34}) \times 10^{38}$ \\ 
2 & $246 \pm 22$ & $3.75 \times 10^{21}$ & 0.21 & 19.6/15 &
$(3.29\pm^{0.45}) \times 10^{38}$ \\
3 & $198 \pm 21$ & $4.38 \times 10^{20}$ & 0.45 & 22.7/14 &
$(3.52 \pm^{0.70}) \times 10^{37}$ \\  
4 & $440 \pm 26$ & $8.34 \times 10^{20}$ & 0.33, 1.03 & 36.3/20 &
$(1.09 \pm^{0.27}_{0.29}) \times 10^{38}$ \\ 
5 & $1268 \pm 39$ & $1.28 \times 10^{21}$ & 0.27, 0.67 & 55.6/45 &
$(2.98 \pm^{0.79}_{0.71}) \times 10^{38}$ \\ 
6 & $668 \pm 30$ & $1.08 \times 10^{21}$ & 0.39, 1.13 & 34.8/29 &
$(1.40 \pm^{0.27}_{0.38}) \times 10^{38}$ \\ 
7 & $45 \pm 17$ & $4.69 \times 10^{21}$ & 0.27 & 4.20/7 & $(5.44
\pm^{47.29}) \times 10^{37}$ \\ 
8 & $442 \pm 25$ & $1.50 \times 10^{21}$ & 0.23, 0.58 & 19.5/20 &
$(1.33 \pm^{0.40}_{0.45}) \times 10^{38}$ \\  
9 & $294 \pm 23$ & $1.42 \times 10^{21}$ & 0.29, 1.93 & 11.96/17 &
$(8.94 \pm^{2.68}_{2.62}) \times 10^{37}$ \\ \hline 
\end{tabular}
\end{center}
\end{table*}

\begin{table*}
\begin{center}
\caption{The $(m-s)/(m+s)$ hardness ratio for each of the 9 regions. Column 
1 lists the region numbers as shown on Fig.~\ref{9}, columns 2 and 3
contain the count rates in the soft and medium bands respectively and
column 4 gives the calculated hardness ratios. The softer emission is shown
by the regions with more negative values.}
\begin{tabular}{|c|c|c|c|} \hline
Region & Count rate in & Count rate in & \underline{(m - s)} \\
 & Soft band (counts/s$^{-1}$) & Medium band (counts/s$^{-1}$) & (m + s) \\ \hline
1 & $(5.33 \pm 0.57) \times 10^{-3}$ & $(8.81 \pm 0.77) \times 10^{-3}$ &
$0.246 \pm 0.070$ \\
2 & $(7.21 \pm 0.67) \times 10^{-3}$  & $(2.78 \pm 0.49) \times 10^{-3}$ &
$-0.443 \pm 0.091$ \\
3 & $(5.39 \pm 0.59) \times 10^{-3}$ & $(3.26 \pm 0.51) \times 10^{-3}$ &
$-0.246 \pm 0.093$ \\
4 & $(11.0 \pm 0.80) \times 10^{-3}$ & $(7.81 \pm 0.70) \times 10^{-3}$ &
$-0.170 \pm 0.057$ \\
5 & $(31.5 \pm 1.20) \times 10^{-3}$ & $(22.4 \pm 1.10) \times 10^{-3}$ &
$-0.169 \pm 0.031$ \\
6 & $(14.4 \pm 0.90) \times 10^{-3}$ & $(13.9 \pm 0.90) \times 10^{-3}$ &
$-0.018 \pm 0.046$ \\
7 & $(1.29 \pm 0.37) \times 10^{-3}$ & $(0.01 \pm 0.38) \times 10^{-3}$ &
$-0.985 \pm 0.570$ \\
8 & $(10.7 \pm 0.80) \times 10^{-3}$ & $(7.23 \pm 0.67) \times 10^{-3}$ &
$-0.194 \pm 0.059$ \\
9 & $(5.33 \pm 0.58) \times 10^{-3}$ & $(6.43 \pm 0.64) \times 10^{-3}$ &
$0.094 \pm 0.074$ \\ \hline
\end{tabular}
\end{center}
\end{table*}

These results are not very conclusive but do suggest a reduced column
density in a region running from north to south across the centre of the
galaxy and also across the northern edge and into the North-East of the
galaxy. The temperature of the softer thermal components seem to be highest
in the North of the galaxy and in an arc running south across the centre of
the galaxy and into the South-East. The hardness ratios suggest that
generally, the emission is slightly higher in the $0.3 - 0.8$~keV energy
band than in the $0.8 - 2.0$~keV band. Looking at the soft, medium and hard
energy band images ($0.3 - 0.8$~keV, $0.8 - 2.0$~keV and $2.0-8.0$~keV
respectively), a difference can be seen between the spatial distribution of
the extended diffuse emission in these 3 bands, (Fig.~\ref{softmed}). These
bands were originally chosen to have roughly equal counts in the 3 bands
and they highlight the relative contributions made by the diffuse emission
and point sources to the different bands. Surprisingly, the $0.3 - 0.8$~keV
band image shows the diffuse emission to be extending to the west of the
galaxy into regions where the spectral fits gave some of the highest column
densities. Such high absorption should reduce the soft thermal emission and
suggests that the single temperature fits in these regions are not the best
models. It is worth noting here that when two temperature models were
fitted to regions 1, 2 and 3, lower column densities were obtained but none
of the results were well constrained.

\begin{figure*}
\vspace{12.0cm}
\includegraphics{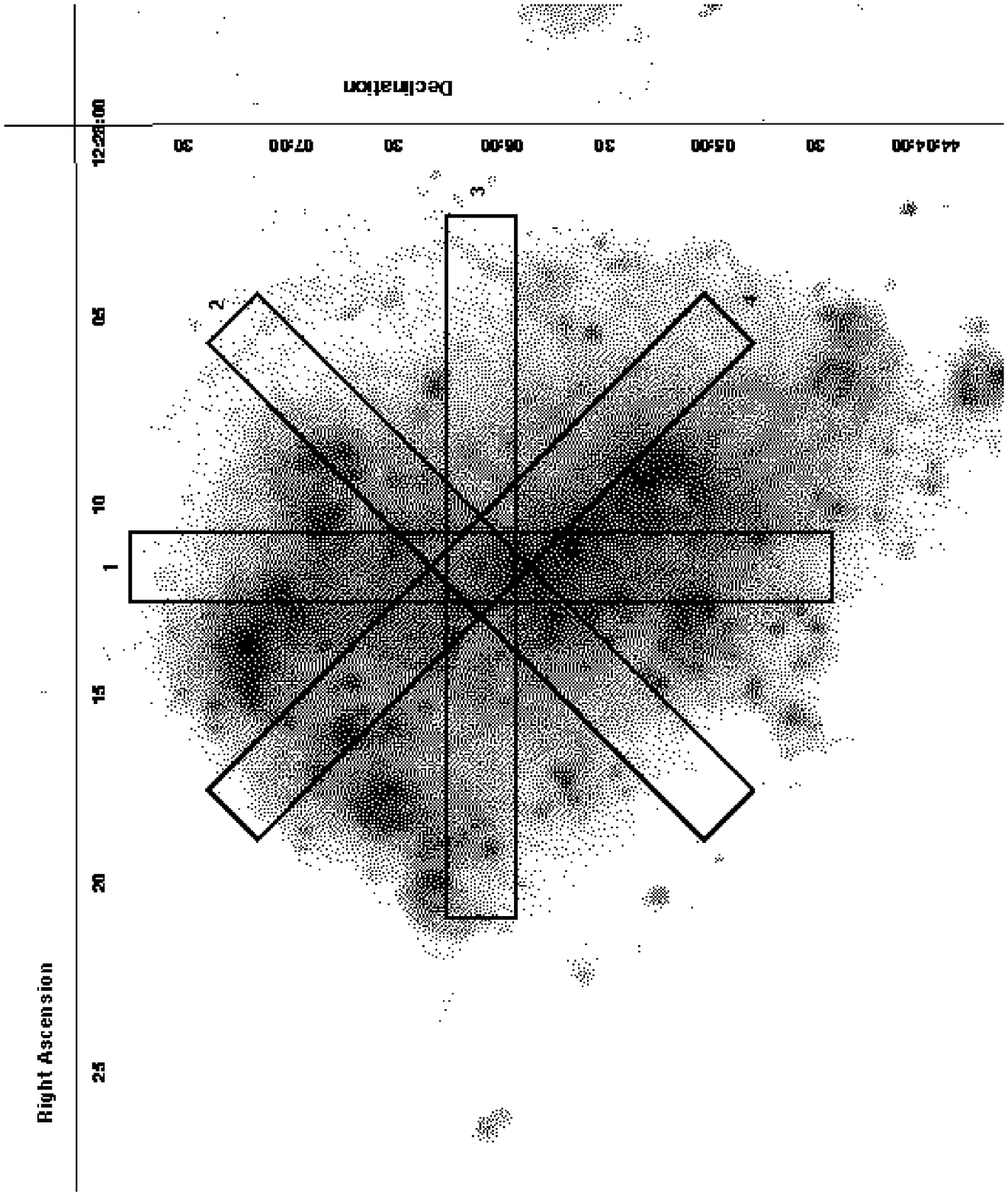}
\caption{H$\alpha$ image of NGC~4449 with the positions of the four slices
overlaid. Each slice is centred on $\alpha = 12^{h} 28^{m} 11.7^{s}$ and
$\delta = +44^{\circ} 06{'} 04.7{''}$, has a length of $200{''}$ and a
width of $20{''}$. Slice 1 runs from North to South, slice 2 from
North-west to South-east, slice 3 from West to East and slice 4 from
South-west to North-east. The resulting X-ray and H$\alpha$ profiles from
these four slices are shown in Fig.~\ref{slices}.}
\label{slice_pos}
\end{figure*}

\begin{figure*}
\vspace{13.0cm}
\includegraphics{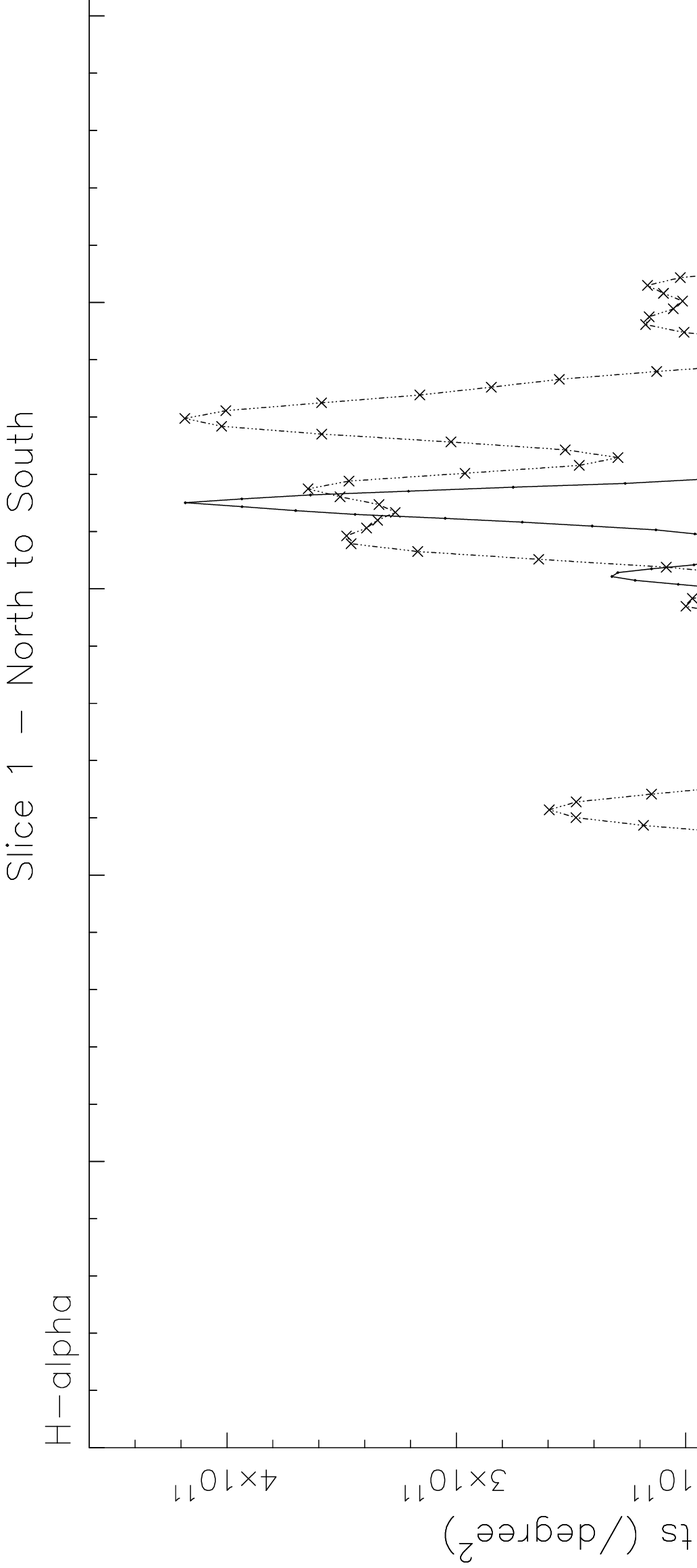}
\includegraphics{MC733_nwse_fig12.ps}
\includegraphics{MC733_we_fig12.ps}
\includegraphics{MC733_swne_fig12.ps}
\caption{X-ray surface brightness  slices (crosses and dashed curves)
overlaid on the H$\alpha$ surface brightness slices (solid curves). The
y-axes on the left of each panel gives the counts/degree$^{2}$ for the
H$\alpha$ slices while the ones to the right are the values for the X-ray
slices. Although these axes are on very different scales they do allow a
comparison to be made of the spatial distribution of X-ray and H$\alpha$
peaks along each of the slices. All the slices are centred on $\alpha =
12^{h} 28^{m} 11.7^{s}$ and $\delta = +44^{\circ} 06{'} 04.7{''}$, located
$0.028^{\circ}$ along the slice, have a length of $200{''}$ and a width of
$20{''}$. This position coincides with the central trough in the diffuse
X-ray emission indicated by the small central contour labelled 2 at that
position in Fig.~\ref{hax}.}
\label{slices}
\end{figure*}

To obtain a clearer picture of the distribution of peaks in the X-ray
emission, surface brightness slices were taken across the galaxy in 4
directions (N to S, NW to SE, W to E and SW to NE), centred on the trough
of the X-ray emission in the centre of the galaxy at $\alpha = 12^{h}
28^{m} 11.7^{s}$, $\delta = +44^{\circ} 06{'} 04.6{''}$ with a length of
$200{''}$ and a width of $20{''}$. The positions of these slices are
overlaid on the H$\alpha$ image of NGC~4449 (kindly supplied by Deidre
A. Hunter) shown in Fig.~\ref{slice_pos}. In addition surface brightness
profiles were also taken from the H$\alpha$ image using the same
slices. The resulting profiles from these slices are shown in
Fig.~\ref{slices}, with those from the diffuse X-ray emission, shown by the
crosses and dashed curves, overlaid on those from the H$\alpha$ emission
(solid curves). Although the X-ray data has been point source subtracted,
the H$\alpha$ data has not. There is very little correlation between the
positions of the resolved X-ray point sources and the numerous HII regions
seen in the H$\alpha$ data and so the slices are being used to look for
association of the diffuse X-ray emission with the star-forming
regions. See Section 5.1.2 for further discussion.

\section{Discussion}

It is apparent from Fig.~\ref{softmed} that the hardest ($> 2$~keV)
fraction of the diffuse emission is confined to the central region of the
galaxy along a ridge running from NNE to the centre. This is in the same
direction as the orientation of the major axis of the galaxy. The medium
($0.8 - 2.0$~keV) diffuse emission is less extended than the soft ($<
0.8$~keV) emission with the most intense regions of emission in both these
bands being coincident with the regions where the highest density of star
clusters is seen (as identified by Gelatt et al. 2001), suggesting that
this emission is associated with the increased star-formation occurring in
these regions. The less intense bulging of the emission seen to the ESE and
WNW could be indicative of a bipolar outflow along the minor axis of the
galaxy as would be expected from standard superbubble models (Weaver et
al. 1977).  The higher column densities seen in Table~6 and Fig~\ref{maps}
for the regions to the SE, SW and W of the galaxy would suggest that these
areas lie behind more absorbing material present in the disk of the galaxy
and that the eastern and northern edges of the galaxy are tilted towards us
as shown in the cartoon representation of the galaxy's morphology in
Fig.~\ref{morph}. Allowing for NGC~4449's position angle, this would mean
the minor axis lies inclined to our line-of-sight (tilted both to the N and
W) consistent with NGC~4449 being a Magellanic Irregular that is inclined
at $56.2^{\circ}$ to our line-of-sight. The lower column density of
region 3, in particular, would suggest that the X-ray emission in that
region comes from material lying above the galactic disk.

\begin{figure*}
\vspace{21.0cm}
\includegraphics{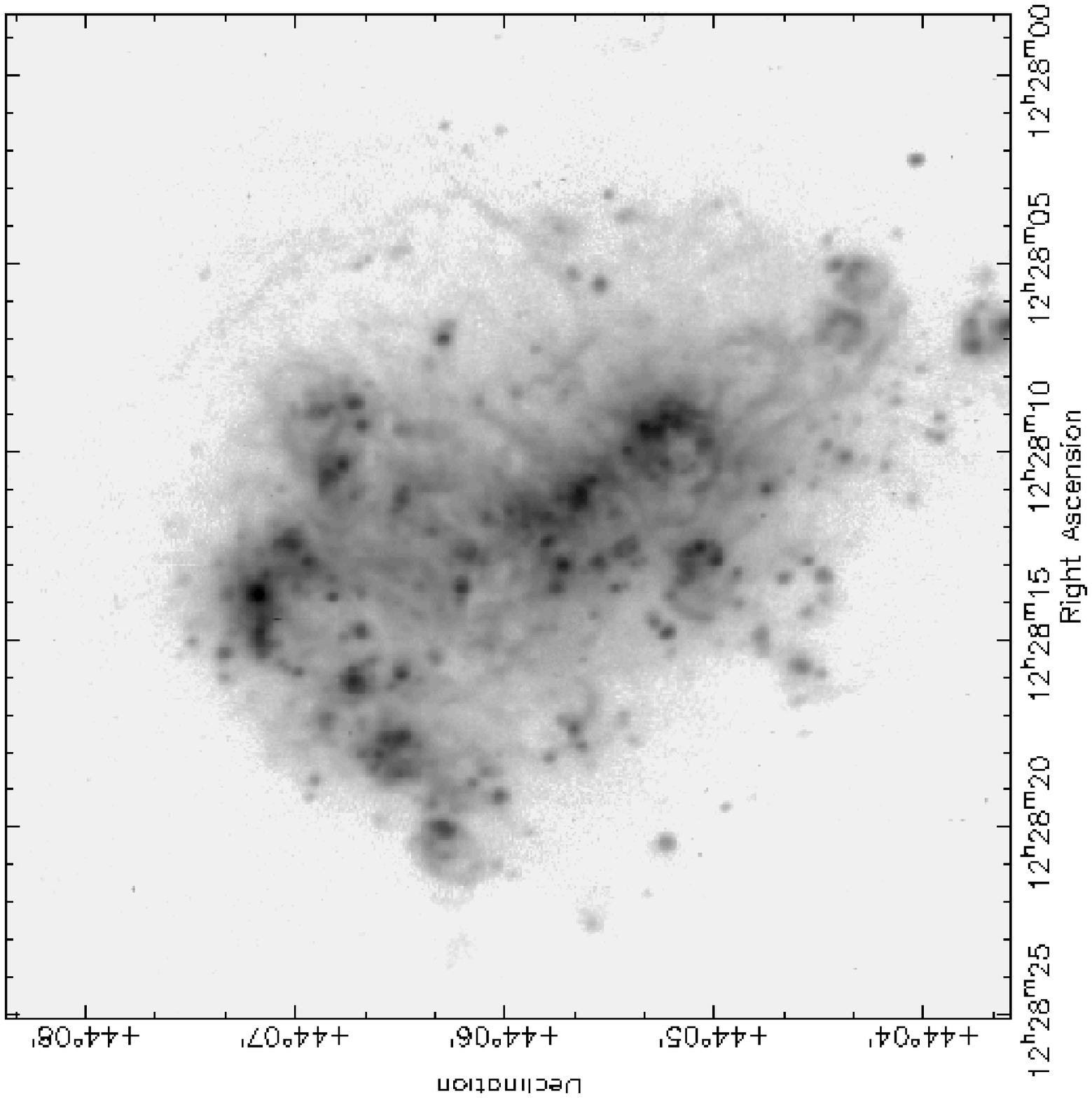}
\includegraphics{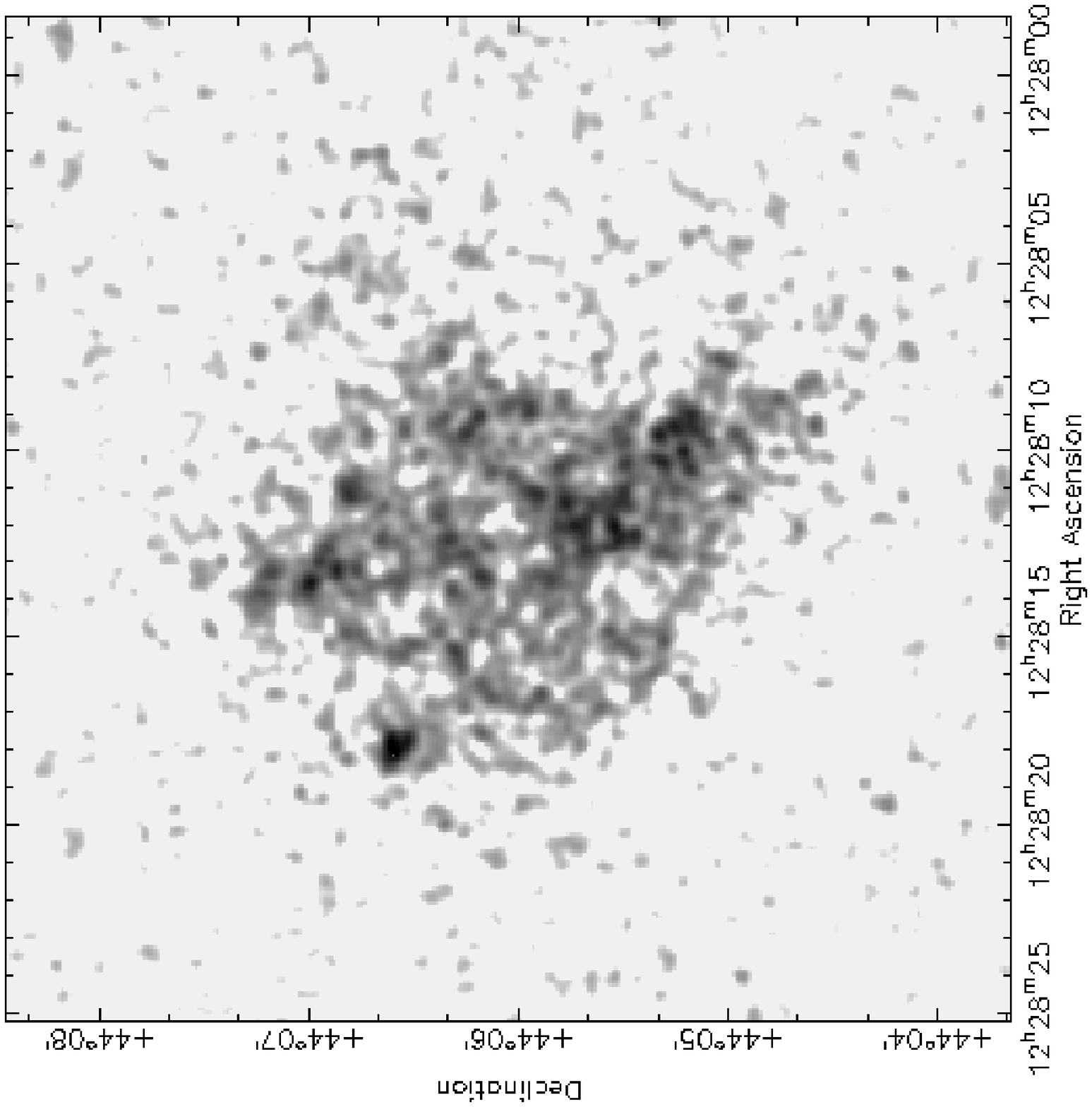}
\includegraphics{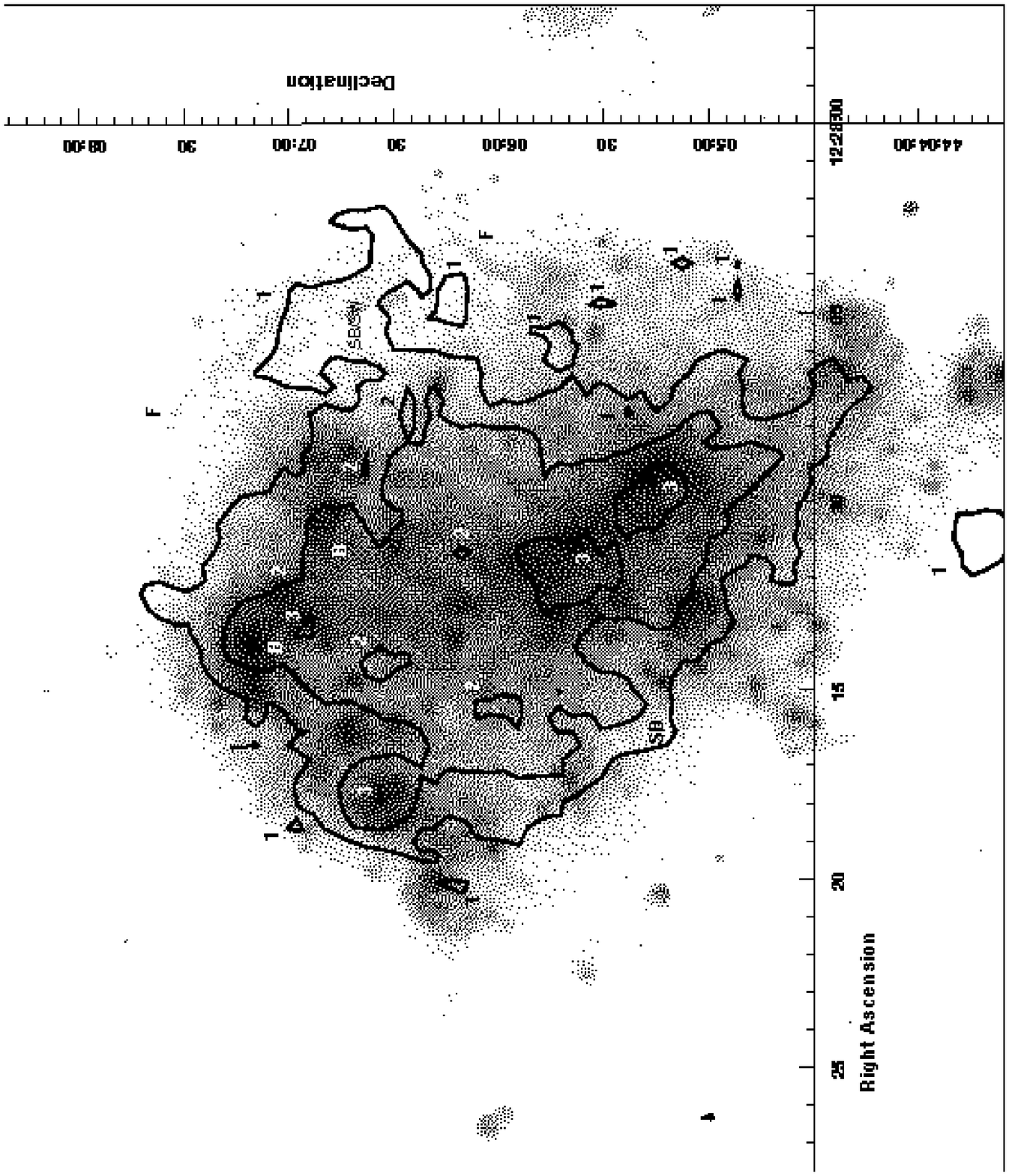}
\caption{Top left: H$\alpha$ grey-scale image of NGC~4449. Top right:
Point source and background subtracted diffuse X-ray grey-scale image of
NGC~4449 on the same spatial scale as the H$\alpha$ image and with the same
smoothing as in Fig.~\ref{9}. Bottom: X-ray contours from the image shown
top right overlaid on the H$\alpha$ image of NGC~4449. The X-ray contours
labelled 1 are at a flux density of $0.65 \times
10^{-13}$~erg~s$^{-1}$~cm$^{-2}$~arcmin$^{-2}$. Those labelled 2 are at a
flux density of $1.3 \times 10^{-13}$~erg~s$^{-1}$~cm$^{-2}$~arcmin$^{-2}$
and those labelled 3 are at $2.6 \times
10^{-13}$~erg~s$^{-1}$~cm$^{-2}$~arcmin$^{-2}$.  The background flux
density level is $\sim 0.25 \times
10^{-13}$~erg~s$^{-1}$~cm$^{-2}$~arcmin$^{-2}$. Also labelled are the
filaments (F), bubbles (B), super-bubbles (SB) and super-wind (SW)
discussed in the text.}
\label{hax}
\end{figure*}

\subsection{Comparison with Observations at Other Wavelengths}

\subsubsection{Star Clusters}

The galaxy contains many star clusters spread through-out the $D_{25}$
ellipse and so star-formation is not confined to a single well-defined
region within the galaxy. The highest density of star clusters identified
by Gelatt et al. (2001) is coincident with the peak seen in the diffuse
X-ray emission at $\alpha = 12^{h} 28^{m} 11.75^{s}$ and $\delta =
+44^{\circ} 05{'} 36.1{''}$. (The largest of the contours labelled 3 in the
lower panel of Fig.~\ref{hax}.) The peak in the diffuse emission to the SW
of this at $\alpha = 12^{h} 28^{m} 09.65^{s}$ and $\delta = +44^{\circ}
05{'} 11.5{''}$ also has a large number of star clusters associated with
it. It is likely that multiple superbubbles are being blown by the combined
action of stellar winds and SN explosions from the individual stars within
these star clusters, leading to the increased diffuse X-ray emission from
hot, shock-heated gas in these regions. This is supported by the complex
morphology seen in the patterns of peaks in the X-ray emission shown in
Fig.~\ref{slices}.

\subsubsection{Comparison with H$\alpha$ Emission}

Fig.~\ref{hax} shows X-ray contours overlaid on an H$\alpha$ map of
NGC~4449 as well as the two separate images. This shows that the X-ray and
H$\alpha$ morphologies in the main body of the galaxy follow each other
closely. The X-ray emission is also seen to extend out to the NW into what
appears to be a hole in the H$\alpha$ emission, bounded by filaments,
labelled F in the lower panel. This could be a wind-blown super-bubble or
the escape of hot gas from a ruptured bubble in the form of a super-wind,
labelled SB/SW. Ultra-violet observations (Hill et al., 1998) have
identified regions along the north-eastern edge of this hole which show
sequential star-formation and this activity may be linked to the expansion
of a super-bubble. As the bubble expands, it sweeps up and shock heats the
ISM, leading to the formation of areas of increased density which can
trigger star-formation. In addition, the observed filamentary structure
seen in H$\alpha$ could be material from a ruptured bubble or entrained ISM
material caught up in an outflow.  A similar, if less pronounced effect
also seems to be occurring to the SE, labelled SB.  The extent of both
emissions matches well, as can be seen by comparing the X-ray and H$\alpha$
surface brightness slices shown for the same regions in
Fig.~\ref{slices}. The only exception seems to occur to the West of the
galaxy where the diffuse X-ray emission appears to extend beyond the region
of high H$\alpha$ intensity.

For the N to S and SW to NE slices, the overall profiles of the peaks are
similar and they occupy a similar extent. The peaks of X-ray emission, in
the northern half of these slices, tend to fall inside the peaks of
H$\alpha$ emission (i.e. they are slightly closer to the centre of the
galaxy), with the intensity of the X-ray emission, in the N, being greatest
close to the H$\alpha$ peaks. This is the type of behaviour discussed by
Strickland et al. (2002) in model 5 of the connection between X-ray and
H$\alpha$ emission, where the H$\alpha$ emission is attributed to a
swept-up shell that has cooled, and surrounds a hot bubble of
SN-ejecta. These potential bubbles in the N of the galaxy, labelled B in
the lower panel of Fig.~\ref{hax}, have diameters $\leq 1$~kpc and the most
northern of the two does appear to be surrounded by an arc of increased
intensity in the H$\alpha$ emission, most clearly visible in the upper left
panel of Fig.~\ref{hax}.  The increased intensity of the X-ray emission
behind the H$\alpha$ peaks would be due to conductive evaporation of the
swept-up material, which increases the density and X-ray emissivity of the
region where the hot gas and evaporated material mixes (Weaver et
al. 1977). These two slices lie closest to the major axis of the galaxy and
cut across the regions containing the highest densities of star clusters in
the southern half of the galaxy. The association of the peaks is less clear
on the south side of both of these slices and this may well be a result of
the overlapping and merging of many wind-blown bubbles from the large
number of star clusters found there.

The slices running from NW to SE and W to E lie closest to the minor axis
of the galaxy and cut across the bulging of the diffuse X-ray emission. The
slices show what appear to be cavities in the H$\alpha$ emission to both
the NW and the E which are filled with X-ray emission, again suggesting the
presence of wind-blown super-bubbles, while to the SE and W, the X-ray
emission seems to lie outside the H$\alpha$ emission, more suggestive of
ruptured super-bubbles, particularly to the west as seen by the
distribution of X-ray contours, labelled 1, in the lower panel of
Fig.~\ref{hax}.

\subsubsection{The Large HI Halo and HI Clouds}

The HI halo of NGC~4449 extends out to a radius of $\sim40$~kpc (Bajaja et
al. 1994) and its outer structure takes the form of enormous filaments and
clouds (Hunter et al. 1998). The VLA observations of Hunter et al. (1998;
1999), show that the inner regions consist of a concentration of gas
centred on the optical galaxy that takes the form of seven large HI
complexes embedded in a lower density background, with the overall outline
of the morphology bulging to the ESE and WNW as seen in the diffuse X-ray
emission.  Fig. 12 of Hunter et al. (1999), shows that the HI in the inner
regions extends beyond the H$\alpha$ filament seen to the WNW in
Fig.~\ref{hax} and beyond the bulge in the diffuse X-ray emission to the
ESE. This could be indicative of the swept up, compressed ambient
interstellar medium lying outside the shells of two expanding
super-bubbles. The regions of increased X-ray emission seen in the main
body of the galaxy as well as lying close to regions of increased H$\alpha$
emission also occupy areas where three of the HI clouds lie. This is
suggestive of the increased HI density in these regions being caused by
outflows from star clusters sweeping-up and compressing the ISM. The
overall extent of the inner HI concentration is $\sim 9$~kpc and it is
embedded in an elliptical region of lower concentration with a major axis
of $\sim 40$~kpc from which streamers extend. By comparison, the extent of
the diffuse X-ray emitting region is $\sim 2.4$~kpc along the major axis of
the optical galaxy and $\sim 1.6$~kpc in an approximate ESE -- WNW
direction. The optical extent of the galaxy as measured by the $D_{25}$
ellipse is $\sim 4.35$~kpc by $\sim 3.35$~kpc. Counter rotation is observed
between the HI gas within the central region and that outside with the
rotation velocity of the inner gas being $\sim 18$~km~s$^{-1}$ and that of
the streamers seen in the halo $\sim 110$~km~s$^{-1}$. These two velocities
and the distribution of the HI gas have implications for the potential of
the hot gas, contained within any super-bubbles to escape. Modelling the
potential of the galaxy as a simple spherically-symmetric, truncated,
isothermal potential (Binney \& Tremaine 1987) then the escape velocity at
a distance $r$ from the centre of the galaxy is given by:

\begin{equation}
v_{esc}(r) = 2^{1/2}v_{rot}[1 + ln(r_{t}/r)]^{1/2}\;\;\;{\rm km~s}^{-1}
\end{equation}

\noindent where $v_{rot}$ is the maximum rotation velocity of the galaxy in
km~s$^{-1}$ and $r_{t}$ is the radius at which the potential is
truncated. If the potential is truncated at the edge of the inner HI
concentration at $r_{t}\sim 4.5$~kpc then the escape velocity for the hot
X-ray emitting gas at $1.6$~kpc from the centre, assuming $v_{rot}\sim
18$~km~s$^{-1}$ would be $\sim 36$~km~s$^{-1}$. This is an upper limit as
the value will decrease when the extent of the hot gas is deprojected. A
similar calculation for the potential truncated at the edge of the HI halo
at $\sim 40$~kpc, with $v_{rot}\sim 110$~km~s$^{-1}$ gives an upper value
for $v_{esc}\sim 320$~km~s$^{-1}$. In the absence of radiative cooling, hot
gas can escape the galaxy's potential if its temperature is greater than:

\begin{equation}
T_{esc} = 1.5 \times 10^{5} (v_{100})^{2}\;\;\;{\rm K}
\end{equation}

\noindent where $T_{esc}$ is the temperature required for the gas to
exceed the galaxy's escape velocity and $v_{100}$ is the escape velocity in
units of $100$~km~s$^{-1}$ (Martin 1999). The two escape velocities
determined above have corresponding escape temperatures of $1.94 \times
10^{4}$~K and $1.54 \times 10^{6}$~K respectively. From the gas parameters
given in Table~5 both the soft and medium components of the hot gas have
temperatures in excess of these values. Eventual escape of the
metal-enriched SN ejecta contained within the super-bubbles seems possible
for NGC~4449 but the distribution of the gas in the galaxy's extended halo
could be as important as gravity in controlling whether the hot gas escapes
or not. The presence of streamers of HI rather than an homogeneous
distribution of material can facilitate the escape of hot gas through the
gaps between the streamers.

\subsubsection{Radio-X-ray Agreement}
The 5~GHz VLA observation (Bignell \& Seaquist 1983) identifies the known
SNR (source 15) in NGC~4449 with the most intense peak of the radio
emission. The next 3 most intense peaks on this radio observation also lie
close to sources in our data (sources 10, 18 and 23) that have all been
flagged as possible SNRs, although none convincingly. In addition, these
radio peaks are coincident with the three brightest peaks of emission on
the H$\alpha$ map, where intense star-formation is occurring and where the
presence of SNRs in star clusters is to be expected.

\subsection{Implications of the Calculated Gas Parameters on the Fate of NGC~4449}

The presence of what appears to be a developing super-wind in this dwarf
galaxy can have dramatic effects on its evolution. The lower gravitational
potential of dwarfs makes the possibility of losing newly synthesized
metals and swept-up ISM material more likely. As such it is worth
considering the impact that the western super-bubble/super-wind might have
on NGC~4449. The position of the super-bubble suggests that its origins may
lie in 3 OB associations identified by Hill et al. (1994), that lie in the
SW region of the main body of the galaxy where the highest density of star
clusters is seen. The ages of these 3 OB associations were found to be,
5.9, 6.2 and 6.7~Myr while their corresponding masses were $1.6 \times
10^{5}$, $6.3 \times 10^{5}$ and $2.5 \times 10^{6}$~$M_{\odot}$. The
extent that the diffuse emission extends from the position of these OB
associations is $\sim 1.7{'}$ or 1.4~kpc for our assumed distance down to a
flux density level of $0.65 \times
10^{-13}$~erg~s$^{-1}$~cm$^{-2}$~arcmin$^{-2}$. Assuming an average age for
the OB associations of 6.3~Myr then to have travelled this distance the
expansion velocity of the bubble has to be $\sim 220$~km~s$^{-1}$. This
will be an underestimate since the deprojected distance will be greater
than that assumed and also the presence of a density gradient in the halo
above the disk of the galaxy will lead to acceleration of the
super-wind. Further, from the calculated gas temperature of the soft
component of the emission given in Table 5, a simple energy conservation
analysis would give the average speed of a particle to be $\sim
280$~km~s$^{-1}$, which is in good agreement. At these sorts of velocities,
it would take $\sim 200$~Myr for the super-wind to reach a distance of
40~kpc and so escape the HI halo. This is of the same order as the
radiative cooling time for the soft component of the diffuse emission and
again suggests that NGC~4449's huge HI halo may allow it to retain its
newly synthesized metals and ISM. The likelihood that the HI halo will
prevent blowout is also found when the criterion for blow-out defined by
Mac Low \& McCray (1988) is applied to NGC~4449. This is based on the
parameter $\Lambda$, defined as the dimensionless rate of kinetic energy
injection, given by:

\begin{equation}
\Lambda = 10^{4} L_{mech,41} H^{-2}_{kpc} P^{-3/2}_{4} n^{1/2}_{0}
\end{equation}

\noindent where $L_{mech,41}$ is the mechanical energy luminosity in units 
of $10^{41}$~erg~s$^{-1}$, $H_{kpc}$ is the galaxy scale-height in kpc,
$P_{4}$ is the initial pressure of the ISM in units of $P/k =
10^{4}$~K~cm$^{-3}$ and $n_{0}$ its initial density. For blow-out to occur,
the condition $\Lambda > 100$ has to be satisfied. For the whole of the
X-ray emitting gas, the thermal energy content given in Table 5 is $2.46
\times 10^{55}$~erg. Assuming that this is the result of complete
thermalization of the kinetic energy of the stellar-winds and SN ejecta,
occurring over the lifetime of the OB associations, then $L_{mech}$ for the
whole galaxy is $\sim 1.2 \times 10^{41}$~erg~s$^{-1}$ (a similar estimate
to the $3 \times 10^{41}$~erg~s$^{-1}$ obtained by Della Ceca et al, 1997,
from the Leitherer \& Heckman, 1995, starburst models). The actual
thermalization will be less than complete, but as argued by Strickland \&
Stevens (2000), will most likely be between 10 and 100 \%, meaning that the
$L_{mech}$ value above could be greater by up to an order of magnitude.
Assuming the initial ISM pressure was similar to that of our Galaxy then
$P_{4} \sim 1$ and comparison to the gas pressures for the hot gas given in
Table 5 show that they are an order of magnitude higher which is to be
expected for an adiabatically expanding over-pressured bubble. As a further
assumption, the hot gas will be at a lower density than the initial ambient
density of the ISM and an order of magnitude change will be assumed from
the average value in Table 5. The result of using these figures in the
above equation with $\Lambda = 100$ means that the scale-height for
NGC~4449 has to be $\leq 9.5$~kpc which is somewhat less than the size of
the HI halo. In addition this figure will be an overestimate since
$L_{mech}$ was estimated from the energy content of the whole volume
assumed to be occupied by the diffuse emission. The volume of the western
superbubble is only $\sim 10\%$ of the total volume and if the ambient ISM
density is assumed to be closer to that of the hot gas then again the
scale-height will decrease. If the halo has holes in it, as suggested by
the presence of streamers rather than an homogeneous distribution, then
escape is more likely.

If escape occurs then the amount of mass lost may affect NGC~4449's ability
to continue star-formation. Assuming the super-bubble has around a tenth of
the volume of the diffuse emission and as such around a tenth of the mass
given in Table 5 then over $6.3$~Myr, the average mass injection rate will
have been 0.14~$M_{\odot}$~yr$^{-1}$, a figure which is similar to that
predicted for Mrk~33 (Summers et al. 2001), another dwarf galaxy that may
be developing a super-wind. If this rate of mass injection was maintained
while the super-bubble/super-wind expanded to the edge of the HI halo over
a time-span of $\sim 200$~Myr then the total amount of mass it would
contain would be $\sim 2.5 \times 10^{7}$~$M_{\odot}$. This value is
greater than the mass contained within the 3 OB associations ($\sim 3.3
\times 10^{6}$~$M_{\odot}$) and the time-scale is also in excess of the
life-times of the massive stars found there, so unless a large amount of
mass-loading from ISM material is to occur this would be an over-estimate
of the mass that the super-bubble/super-wind would contain on
escape. Compared to the total mass of NGC~4449 of $\sim 4 \times
10^{10}$~$M_{\odot}$ this is only about $0.1\%$ of the galaxy's total
mass. As the time-scale required is in excess of the life-times of the
massive stars then the energy-injection rate would also decrease before
escape was attained, suggesting that the bubble expansion is likely to
stall. Hence it seems unlikely that the development of this
super-bubble/super-wind in NGC~4449 will have a great deal of effect on the
galaxy's ability to both retain its products of star-formation and continue
star-formation in the future.

\begin{figure*}
\vspace{10.0cm}
\includegraphics{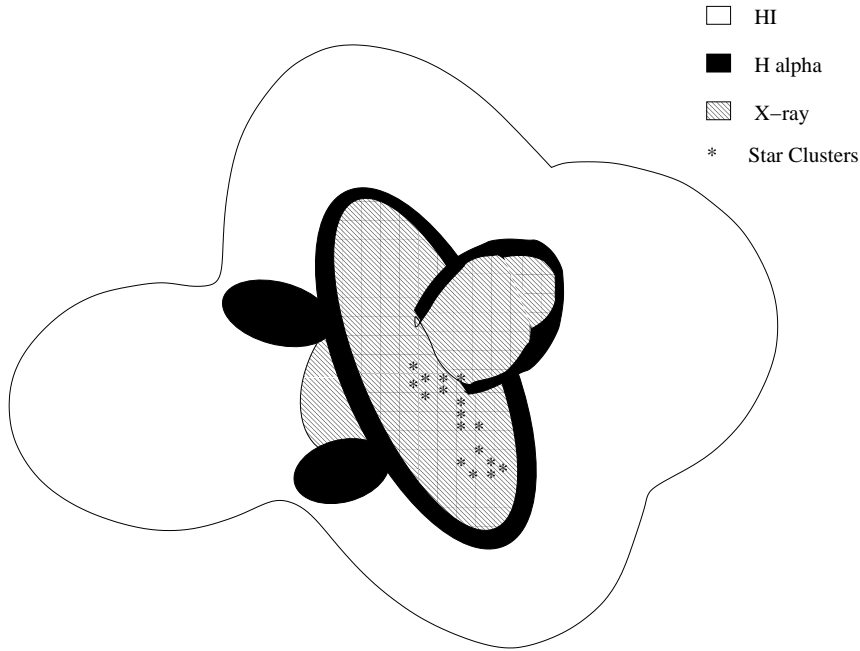}
\caption{Cartoon representation of the relationship between the X-ray,
H$\alpha$ and HI morphologies of NGC~4449 and the approximate positions of
the starclusters in the densest groups (from Gelatt et al. 2001). The
extent of the X-ray and H$\alpha$ emissions were determined from a
comparison of matched images of the X-ray data and the H$\alpha$ image. The
extent of the HI shown is estimated from Fig.~12 of Hunter et al. (1998)
which shows HI contours overlaid on an image of the same H$\alpha$
data. The structure shown in the X-ray bubble to the NW represents the
extension of the X-ray emission seen in this direction in Fig.~\ref{hax}
that may actually extend beyond the H$\alpha$ filament seen there and as
such may be indicative of the escape of hot gas from a ruptured
super-bubble in the form of a super-wind.}
\label{morph}
\end{figure*}

\subsection{Morphology of NGC~4449}
Fig.~\ref{morph} shows a cartoon of the possible morphology of NGC~4449 and
in particular shows how the diffuse X-ray, H$\alpha$ and HI emissions may
be associated with each other, at the maximum extent of the X-ray and
H$\alpha$ emission, if the galaxy is assumed to have a bi-polar outflow in
the form of two wind-blown super-bubbles along its minor axis.  The stars
represent the approximate positions of the starclusters in the two densest
populated regions, as identified by Gelatt et al. (2001). The structure
shown in the X-ray emission in the NW super-bubble may be indicative of the
bubble rupturing and X-ray emission beginning to extend beyond the
H$\alpha$ emission as is suggested in the X-ray and H$\alpha$ slices of
Fig.~\ref{slices}. A comparison of the NW--SE and W--E slices which run
along the northern and southern edges of the bubble respectively suggest
that the X-ray emission is bounded by the H$\alpha$ emission to the north
but not to the South. A similar picture is seen to the East of the galaxy
where the emission to the East seems to be bounded whilst that to the SE is
unbounded. In the regions of intense star-formation to the S, there are
several peaks of emission in both the X-ray and H$\alpha$, indicative of a
confused morphology where the presence of several wind-blown bubbles is to
be expected and their overlapping both physically and in projection will
make individual bubbles impossible to identify. To the N, the presence of 2
bubbles with diameters $\leq 1$~kpc seems to be indicated in the surface
brightness slices of Fig.~\ref{slices}. The HI outline on Fig.~\ref{morph}
represents the extent of the inner HI cloud complexes and HI ISM down to a
column density of $15\times 10^{20}$~cm$^{-2}$ as shown on Figs.~10 and 12
of Hunter et al. (1999). Comparing these two figures with Fig.~\ref{hax}
above, although obviously different in size, all three types of emission do
appear to have a similar outline to their morphologies.

\section{Summary and Conclusions.}
In summary, we have presented an analysis of the X-ray data obtained from a
30~ks observation, of the Magellanic Irregular, starburst galaxy NGC~4449,
by the {\it CHANDRA} satellite. We find X-ray emission from 32 discrete
point sources in the S3 chip data with 24 of them lying within the optical
extent of the galaxy, as measured by the extent of the $D_{25}$
ellipse. Some of these can be clearly identified as SNRs, others as XRBs
and there are also several SSS. We calculate an age for the previously
identified SNR (source 15) of $270$~yr and a density for the medium into
which it has exploded of being $\leq 200$~cm$^{-3}$. The bright SSS (source
14) is a candidate for being a white dwarf binary system, while the source
exhibiting the highest count rate during the observation (source 27), shows
long term variability and is most likely an HMXB.

The galaxy has a very extended distribution of HI, but the inner
concentration of HI has an overall general morphology which also follows
that of the X-ray emission but on a larger scale. The HI complexes within
the inner region could be the result of outflows from the multiple star
clusters in the starburst of NGC~4449 compressing and driving the ambient
ISM of the galaxy away from the centre, both along the minor axis of the
galaxy and above regions of increased star formation within the disk. The
overall extent of the inner HI concentration is $\sim 9$~kpc, that of
H$\alpha$ is $\sim 4.35 \times 3.35$~kpc and that of the X-ray emission is
$\sim 2.5 \times 1.6$~kpc., with this central region being embedded within
an huge HI halo extending out to $\sim 40$~kpc.

The total X-ray luminosity of NGC~4449 in the $0.3 - 8.0$~keV energy band
is $(2.46\pm^{0.19}_{0.20}) \times 10^{39}$~erg~s$^{-1}$~cm$^{-2}$. Of
this, $\sim 60\%$ is due to resolved point sources and the rest is due to
diffuse X-ray emission. As suggested by this analysis and those of other
starburst galaxies (e.g. NGC~253, Strickland et al. 2002: NGC 1569, Martin
et al. 2002, Della-Ceca et al. 1996: NGC~4449 Della Ceca et al. 1997), the
diffuse X-ray emission is emerging as a multi-phase environment requiring
complex spectral models which the spectral resolution of even current X-ray
telescopes cannot meaningfully constrain. We thus conclude that the diffuse
X-ray emission contains gas at, at least two different temperatures and
evidence for unresolved point sources. The fitted gas temperatures are
$(0.28 \pm 0.01)$~keV and $(0.86 \pm 0.04)$~keV for the soft and medium
components respectively and their respective absorption corrected fluxes in
the $0.3 - 8.0$~keV band are $(6.66\pm^{0.34}_{0.35}) \times
10^{-13}$~erg~s$^{-1}$~cm$^{-2}$ and $(2.20\pm^{0.26}_{0.25}) \times
10^{-13}$~erg~s$^{-1}$~cm$^{-2}$ corresponding to luminosities of $(6.83
\pm^{0.35}_{0.36}) \times 10^{38}$~erg~s$^{-1}$ and
$(2.26\pm^{0.27}_{0.26}) \times 10^{38}$~erg~s$^{-1}$. This diffuse
emission seems to be more heavily absorbed in the SW of the galaxy and
shows a higher temperature in the central and eastern regions consistent
with NGC~4449 being an inclined Magellanic Irregular galaxy, with its minor
axis inclined to the N and W of our line-of-sight. The morphologies of the
X-ray and H$\alpha$ emissions from the galaxy follow each other closely and
the X-ray emission appears to fill cavities in the H$\alpha$ emission to
the NW and E -- highly suggestive of wind-blown super-bubbles extending
along the minor axis of the galaxy. To the SE and W, the X-ray emission
seems to lie outside the H$\alpha$ emission and this may indicate the
developing of a super-wind from the rupture of the super-bubbles. The large
number of peaks in emission seen in both the H$\alpha$ and diffuse X-ray
data indicates a complex morphology within the main body of the
galaxy. Regions containing small, ($\leq 1$~kpc), bubbles of X-ray emission
surrounded by shells of H$\alpha$ emission are seen, that are produced by
the combined action of stellar-winds and SN explosions from massive stars,
shock-heating both the stellar-ejecta and ISM and sweeping it into shells
as the ejecta drives its way out of star clusters.

The hot X-ray emitting gas has a total thermal energy content of $\sim 2.5
\times 10^{55}$~erg and a total mass of $\sim 9.0 \times
10^{6}$~$M_{\odot}$. The super-bubble/super-wind extending to the WNW of
the galaxy occupies around $10\%$ of the total volume occupied by the hot
gas (assuming a filling factor of 1). The origin of this out-flow appears
to be the concentration of star clusters observed in the SW region of the
galaxy disk (Hill et al. 1994). At present, the average energy injection
rate for the whole galaxy is $\sim 1.2 \times 10^{41}$~erg~s$^{-1}$ and the
mass injection rate into the super-bubble is $\sim
0.14$~$M_{\odot}$~yr$^{-1}$. It seems unlikely that this bubble can escape
from the huge HI halo at its current estimated expansion speed of $\sim
220$~km~s$^{-1}$, as the time required is in excess of the life-times of
the massive stars of the star clusters and comparable to the radiative
cooling time of the soft component of the diffuse emission, whilst the
expansion velocity is less than the estimated escape velocity from the HI
halo. However, the current temperatures of both gas components imply that
they are capable of escaping the galaxy's gravitational potential and so
venting metal-enriched, hot gas into the IGM. The crucial factors are the
time for which energy injection can be maintained and the actual
distribution of HI in the halo of NGC~4449. The less homogeneous the
distribution, the more likely the possibility that some of the hot gas can
escape. At current mass injection rates, the total mass loss would be $<
1\%$ of the galaxy's mass and as such would not have a catastrophic effect
on NGC~4449.

\section*{Acknowledgements}
LKS and IRS acknowledge funding from a PPARC studentship and Advanced
Fellowship respectively. DKS is supported by NASA through {\it CHANDRA}
Postdoctoral Fellowship Award Number PF0-10012, issued by the {\it CHANDRA}
X-ray Observatory Center, which is operated by the Smithsonian
Astrophysical Observatory for and on behalf of NASA under contract
NAS8-39073. Our thanks go to Deidre A. Hunter for kindly providing us with
the H$\alpha$ image of NGC~4449.

\end{document}